\documentclass[twocolumn,letterpaper,aps,prc,superscriptaddress,showpacs,floatfix,preprintnumbers,longbibliography]{revtex4-1}
\usepackage{graphicx} 
\usepackage{amsmath, amssymb}
\usepackage[colorlinks=true, linkcolor=blue, citecolor=blue, urlcolor=cyan, bookmarksnumbered]{hyperref}
\usepackage{xspace}

\newcommand{\Tc}{\ensuremath{T_{\mathrm{C}}}\xspace}

\newcommand{\ee}{\ensuremath{e^{+}e^{-}}\xspace}
\newcommand{\uu}{\ensuremath{\mu^{+}\mu^{-}}\xspace}
\newcommand{\muB}{\ensuremath{\mu_{\mathrm{B}}}\xspace}

\newcommand{\ahfl}{${}^{4}_{\bar{\Lambda}}\overline{\hbox{H}}$}

\begin{document}

\title{Selected highlights from STAR experiment}
\affiliation{Key Laboratory of Nuclear Physics and Ion-beam Application (MOE), Institute of Modern Physics, Fudan University, Shanghai 200433, China}
\affiliation{Key Laboratory of Particle Physics and Particle Irradiation (MOE), Institute of Frontier and Interdisciplinary Science, Shandong University, Qingdao 266237, China}
\affiliation{QMRC, Institute of Modern Physics, Chinese Academy of Sciences, Lanzhou, China}
\affiliation{Key Laboratory of Quark \& Lepton Physics (MOE) and Institute of Particle Physics, Central China Normal University, Wuhan 430079, China}
\affiliation{Department of Modern Physics, University of Science and Technology of China, Hefei, China}
\affiliation{Institute of Quantum Matter, South China Normal University, Guangzhou 510006, China}
\affiliation{Department of Engineering Physics, Tsinghua University, Beijing, China}

\author{Jinhui Chen*}\affiliation{Key Laboratory of Nuclear Physics and Ion-beam Application (MOE), Institute of Modern Physics, Fudan University, Shanghai 200433, China}
\author{Zhenyu Chen}\affiliation{Key Laboratory of Particle Physics and Particle Irradiation (MOE), Institute of Frontier and Interdisciplinary Science, Shandong University, Qingdao 266237, China}
\author{Maowu Nie}
\affiliation{Key Laboratory of Particle Physics and Particle Irradiation (MOE), Institute of Frontier and Interdisciplinary Science, Shandong University, Qingdao 266237, China}
\author{Hao Qiu}
\affiliation{QMRC, Institute of Modern Physics, Chinese Academy of Sciences, Lanzhou, China}
\author{Shusu Shi}
\affiliation{Key Laboratory of Quark \& Lepton Physics (MOE) and Institute of Particle Physics, Central China Normal University, Wuhan 430079, China}
\author{Zebo Tang}
\affiliation{Department of Modern Physics, University of Science and Technology of China, Hefei, China}
\author{Qinghua Xu*}
\affiliation{Key Laboratory of Particle Physics and Particle Irradiation (MOE), Institute of Frontier and Interdisciplinary Science, Shandong University, Qingdao 266237, China}
\author{Chi Yang}
\affiliation{Key Laboratory of Particle Physics and Particle Irradiation (MOE), Institute of Frontier and Interdisciplinary Science, Shandong University, Qingdao 266237, China}
\author{Shuai Yang}
\affiliation{Institute of Quantum Matter, South China Normal University, Guangzhou 510006, China}
\author{Zaochen Ye}
\affiliation{Institute of Quantum Matter, South China Normal University, Guangzhou 510006, China}
\author{Li Yi}
\affiliation{Key Laboratory of Particle Physics and Particle Irradiation (MOE), Institute of Frontier and Interdisciplinary Science, Shandong University, Qingdao 266237, China}
\author{Wangmei Zha}
\affiliation{Department of Modern Physics, University of Science and Technology of China, Hefei, China}
\author{Chunjian Zhang}
\affiliation{Key Laboratory of Nuclear Physics and Ion-beam Application (MOE), Institute of Modern Physics, Fudan University, Shanghai 200433, China}
\author{Jinlong Zhang}
\affiliation{Key Laboratory of Particle Physics and Particle Irradiation (MOE), Institute of Frontier and Interdisciplinary Science, Shandong University, Qingdao 266237, China}
\author{Yifei Zhang}
\affiliation{Department of Modern Physics, University of Science and Technology of China, Hefei, China}
\author{Xianglei Zhu}
\affiliation{Department of Engineering Physics, Tsinghua University, Beijing, China}


\begin{abstract}
 In this paper, we review recent highlights in heavy-ion collisions and proton-proton collisions at top energies from STAR experiment  at the Relativistic Heavy Ion Collider (RHIC) with key contributions from Chinese groups, including the Quark-Gluon Plasma (QGP) bulk properties, electromagnetic probes, heavy flavor and jets, antimatter hyper-nucleus, nuclear structure, global polarization, and nucleon spin structure. These data serve as important ingredients in the physics of Quantum Chromodynamics (QCD). 
\end{abstract}
\maketitle

\section{Introduction}
The theory of the strong interaction is Quantum Chromodynamics (QCD), which exhibits asymptotic freedom and long-range ``color confinement"~\cite{Gross:1973id,Politzer:1973fx}.
In 1970’s, T.D. Lee et al. proposed utilizing relativistic heavy ion collisions to create extreme conditions of high temperature and energy density, thereby altering the nature of the vacuum and thus lifting quark confinement to produce a new state of matter – Quark Gluon Plasma (QGP)~\cite{Lee:1974ma,Lee:1974kn}
The hot and dense medium created in relativistic heavy-ion collision mimics the state of the universe a few microseconds after the ‘Big Bang’, evolving from QGP to hadronic matter.
Dedicated high energy heavy ion colliders have been built to investigate the formation of QGP and its properties, including  Relativistic Heavy Ion Collider (RHIC) at Brookhaven Laboratory and Large Hadron Collider at CERN.
Exploring QGP and studying its properties through experimental and theoretical research has been the mainstream of high energy nuclear physics.
Experimental observables including collective flow, jet quenching, strangeness enhancement, etc, from RHIC have shown that a strongly interacting quark matter with extremely low-viscosity near-perfect fluid has been produced in relativistic heavy-ion collision~\cite{STAR:2005gfr, PHENIX:2004vcz,BRAHMS:2004adc,PHOBOS:2004zne}. 

Chinese scientists started experimental research on QGP and relativistic heavy-ion collisions in the early 2000s when several institutions (Central China Normal University, University of Science and Technology of China, Tsinghua University, Shanghai Institute of Applied Physics, Chinese Academy of Sciences, Institute of Modern Physics, Chinese Academy of Sciences, Fudan University, Shandong University, etc.) joined the RHIC-STAR Collaboration~\cite{Chen:2024aom}. In the past decades, the STAR-China group has made outstanding achievements in detector upgrades and physics analyses in the RHIC-STAR experiment. The STAR-China team designed and constructed the STAR time-of-flight spectrometer TOF, muon detector MTD, also made dominant contributions to the detector upgrades on the inner Time Projection Chamber (iTPC), event plane detector, end-cap TOF, forward tracking detector, etc., and played a leading role in the discovery of anti-hypertriton~\cite{STAR:2010gyg}, anti-alpha~\cite{STAR:2011eej}, $\overline{\rm ^4_{\bar{\Lambda}}H}$~\cite{STAR:2023fbc} and measurement of antimatter interaction~\cite{STAR:2015kha,STAR:2019wjm} in the STAR experiment. STAR-China group also played an essential role in studying the QGP property, QCD phase structure, nucleon spin structure at STAR~\cite{Chen:2024aom,Xu:2015bid,Chen:2023mel}. The STAR-China group now expands to 16 institutions including two from Taipei. 

 In this paper, we review recent highlights in studying QGP in heavy-ion collisions at top energies from RHIC-STAR experiments led by STAR-China group, including bulk properties, electromagnetic probes \& ultra-peripheral collisions, heavy flavor \& jets, nuclear structure, and spin physics including global polarization and nucleon spin structure study with polarized proton-proton collision.

\section {Bulk properties of QGP}

One of the central goals of relativistic heavy-ion physics is to determine whether the hot and dense medium created in nucleus--nucleus collisions exhibits collectivity at the partonic level, prior to hadronization.
Collective flow observables, in particular anisotropic flow coefficients such as elliptic flow ($v_2$), provide direct sensitivity to the early-time pressure gradients and transport properties of the medium.
A key physics question is whether the observed collectivity arises predominantly from hadronic interactions after confinement, or is already established in a deconfined QGP phase.
Systematic measurements of flow for hadrons with different masses, quark content, and different collision systems therefore offer a powerful means to disentangle the relevant degrees of freedom and to test the emergence of partonic collectivity.

\subsection{Partonic collectivity}

At the top RHIC energy of $\sqrt{s_{\rm NN}}=200$ GeV, the STAR experiment has performed comprehensive measurements of elliptic flow ($v_2$), direct flow ($v_1$) for a wide spectrum of hadron species, ranging from light hadrons ($\pi$, $K$, $p$) to multi-strange hadrons ($\Lambda$, $\Xi$, $\Omega$, $\phi$) and open-charm mesons ($D^0$)~\cite{STAR:2003wqp, STAR:2007mum, STAR:2010ico, STAR:2015gge, STAR:2017kkh, STAR:2022ncy, STAR:2022tfp, STAR:2023jdd, Shi:2020htw}.
These measurements provide direct evidence that the strongly interacting medium formed in Au+Au collisions exhibits robust collective behavior and nearly ideal hydrodynamic response.

\begin{figure}
\centering
\includegraphics[width=0.48\textwidth]{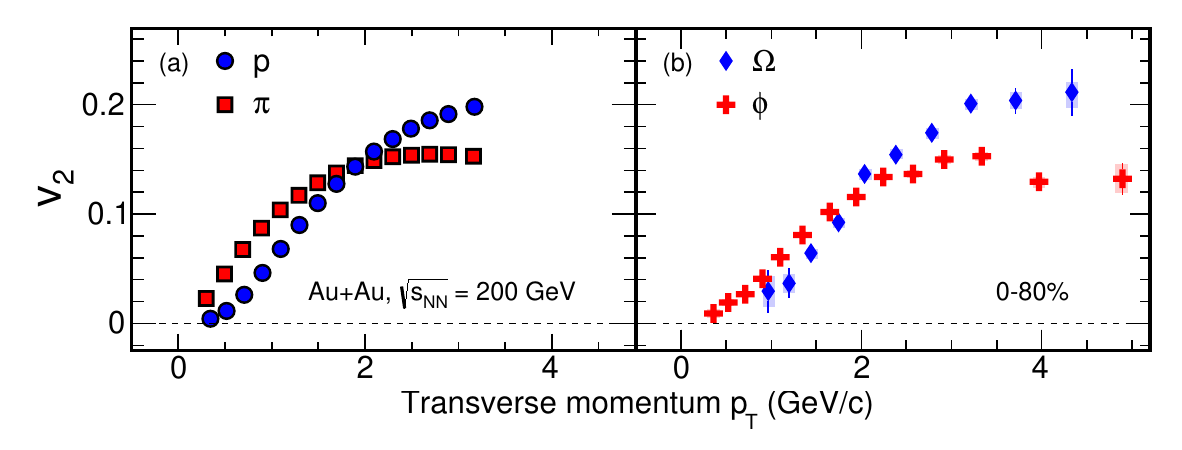}
\caption{The elliptic flow coefficient $v_2$ as a function of transverse momentum $p_{\rm T}$ for (a) light hadrons ($\pi$ and $p$) and (b) multi-strange hadrons ($\phi$ and $\Omega$) in minimum-bias Au+Au collisions at $\sqrt{s_{\rm NN}} = 200$~GeV. Results are adapted from~\cite{STAR:2015gge}.
}\label{phiOmegav2}
\end{figure}

At low transverse momentum ($p_{\mathrm{T}} \lesssim 2$ GeV/$c$), a characteristic mass ordering of $v_2$ is observed, consistent with the expectations of hydrodynamic expansion in which heavier particles acquire smaller flow amplitudes due to their larger inertia.
At intermediate $p_{\mathrm{T}}$ (2–5 GeV/$c$), the $v_2$ values of baryons exceed those of mesons at the same $p_{\mathrm{T}}$, producing a clear baryon–meson splitting.
This behavior is quantitatively described by quark coalescence models, in which hadrons form via the recombination of constituent quarks that already possess collective flow.
When $v_2$ and $p_{\mathrm{T}}$ are scaled by the number of constituent quarks $n_q$, the data for mesons and baryons collapse onto a common curve — the well-known Number-of-Constituent-Quark (NCQ) scaling.
This scaling provides compelling evidence that collectivity develops at the partonic level, before hadronization.

Particularly significant is the observation that multi-strange hadrons such as $\phi$ and $\Omega$ exhibit $v_2$ magnitudes comparable to those of light hadrons, despite their much smaller hadronic interaction cross sections (Fig.~\ref{phiOmegav2}).
Because these particles decouple early from the medium, their substantial elliptic flow indicates that the collective motion is established prior to hadron freeze-out, in the deconfined partonic stage~\cite{Chen:2008vr}.
The $\phi$ meson, composed of $s\bar{s}$, and the $\Omega$ baryon, composed entirely of strange quarks ($sss$), thus provide especially clean probes of early-time QGP dynamics and confirm that strange quarks participate in the collective flow.

\begin{figure}
\centering
\includegraphics[width=0.48\textwidth]{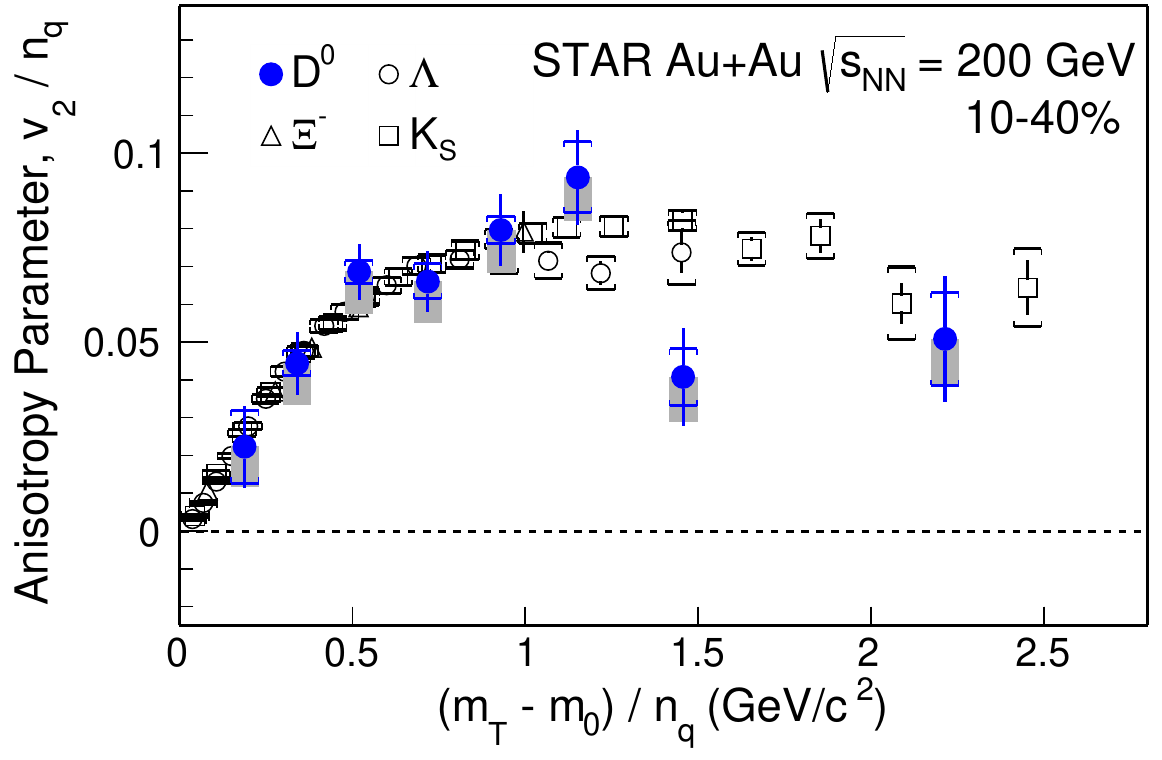}
\caption{The elliptic flow $v_2$, scaled by the number of constituent quarks ($n_q$), as a function of the scaled transverse kinetic energy $(m_T - m_0)/n_q$ for $D^0$, $\Xi^-$, $\Lambda$, and $K_S^0$ in 10–40\% Au+Au collisions at $\sqrt{s_{NN}} = 200$~GeV. Results adapted from~\cite{STAR:2017kkh}. Here, $m_T = \sqrt{p_{\rm T}^2 + m_0^2}$ is the transverse mass.
}\label{D0v2}
\end{figure}

Beyond the strange sector, STAR has reported a significant elliptic flow for open charm hadrons, particularly the $D^0$ meson, which contains a heavy charm quark produced predominantly in the initial hard scatterings~\cite{STAR:2017kkh}.
The observation of sizable $D^0$ $v_2$ values (Fig.~\ref{D0v2}) demonstrates that charm quarks interact strongly with the QGP, acquiring collective flow through frequent scattering with the medium constituents.
When plotted in NCQ-scaled form, the $D^0$ meson follows the same universal trend as light and strange hadrons, implying partial thermalization of charm quarks and reinforcing the picture of partonic collectivity.
These results place strong constraints on the heavy-quark diffusion coefficient and the transport properties of the QGP at RHIC energies.

Elliptic flow measurements at $\sqrt{s_{\rm NN}}=200$ GeV reveal a consistent and unified picture of the QGP as a strongly coupled, nearly perfect fluid in which all quark flavors—light, strange, and even charm—participate in collective expansion. 
The simultaneous observation of large $v_2$ values for multi-strange and open-charm hadrons, together with the success of NCQ scaling, represents one of the most striking signatures of deconfinement and partonic collectivity achieved at RHIC~\cite{STAR:2003wqp, STAR:2007mum, STAR:2015gge, STAR:2017kkh}. 
In particular, the substantial elliptic flow of multi-strange hadrons and $D^0$ mesons highlights the active participation of both strange and heavy quarks in the collective expansion of the medium. 
These results provide strong evidence that collectivity is established at the partonic level prior to hadronization.

Looking forward, precision flow measurements with increased statistics and improved detector capabilities will enable more differential studies of partonic collectivity.
Extending NCQ-scaling tests to higher-order flow coefficients, wider kinematic ranges, and rarer probes will help quantify possible deviations from idealized coalescence expectations and constrain the onset and limits of partonic behavior.
Such measurements are essential for disentangling the roles of hadronization dynamics, initial-state fluctuations, and viscous effects in shaping the observed collective patterns.
Furthermore, future experimental and theoretical efforts focusing on heavy-flavor and multi-strange observables will provide increasingly stringent constraints on the transport properties of the QGP, including the temperature dependence of the shear viscosity and heavy-quark diffusion coefficient.
Combined with complementary measurements across beam energies and collision systems, these studies will deepen our understanding of how partonic collectivity emerges and evolves, offering critical insight into the microscopic structure of strongly interacting QCD matter.

\subsection{Collectivity in small system}

Over the past decade, similar collective signals have been observed in high-multiplicity proton-proton~\cite{Khachatryan:2010gv,Aad:2015gqa,Khachatryan:2015lva,Khachatryan:2016txc,CMS:2020qul}, 
proton-nucleus collisions at the LHC~\cite{CMS:2012qk,alice:2012qe,Aad:2012gla,LHCb:2015coe,CMS:2018loe,CMS:2018duw,CMS:2022bmk,CMS:2024krd}, 
as well as in proton-gold (p+Au), deuteron-gold (d+Au) and helium-3--gold ($^{3}$He+Au) collisions at RHIC~\cite{PHENIX:2014fnc,PHENIX:2015idk,PHENIX:2016cfs,PHENIX:2017djs,PHENIX:2018lia,PHENIX:2021ubk,STAR:2015kak,STAR:2019zaf}. 
These observations raise the question of whether QGP droplets are produced in systems of much smaller size than nucleus-nucleus collisions~\cite{Nagle:2018nvi,Dusling:2015gta}. 
Although hydrodynamic simulations suggest their applicability to   p+Au, d+Au, and $^{3}$He+Au collisions when the system sizes are on the order of the inverse temperature~\cite{Chesler:2016ceu}, the influences of subnucleonic fluctuations on the initial geometry of the collisions have not been fully understood. 
Such fluctuations are introduced by spatially inhomogeneous gluon field distributions inside nucleons~\cite{Welsh:2016siu}. 
Although its impact is negligible by large sized collision systems where the initial geometry is primarily determined by the spatial distribution of nucleons inside the nuclei, 
a key challenge in quantitatively describing the collective phenomena in small sized collisions is the strong dependence on the subnucleonic fluctuations~\cite{Schenke:2021mxx}. 

\begin{figure}
\centering
\includegraphics[width=0.49\textwidth]{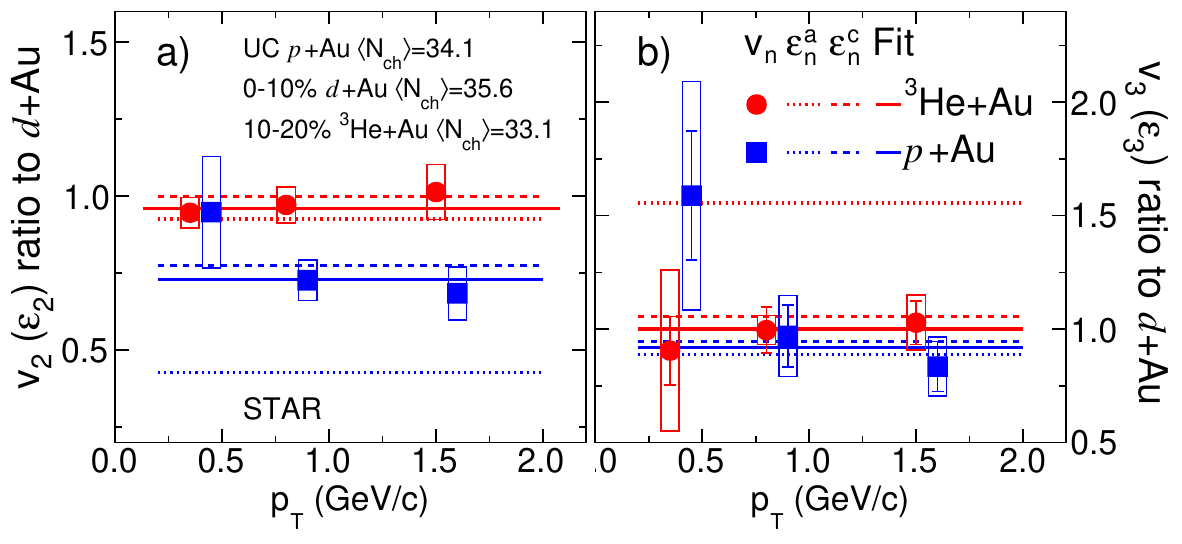}
\caption{
The ratios of $v_2$ (a) and $v_3$ (b) between p+Au ($^{3}$He+Au) and d+Au collisions with a similar number of particles produced in the collisions ($\langle \mathrm{N_{ch}} \rangle$). The solid lines are constant fits to the measurements, and the dashed lines are eccentricity ratios from model calculations with (model $c$) and without (model $a$) subnucleonic fluctuations. The vertical bars indicate statistical uncertainties, while the systematic uncertainties are indicated by the boxes. The figure is taken from~\cite{STAR:2022pfn}.
}
\label{fig:vnsmallsystem} 
\end{figure}

Recent STAR measurements of $v_2$ and $v_3$ as functions of $p_T$ in central p+Au, d+Au, and $^{3}$He+Au collisions provide key information on subnucleonic fluctuations in these small collision systems~\cite{STAR:2022pfn,STAR:2023wmd}. 
Figure~\ref{fig:vnsmallsystem} shows the ratios of $v_2$ and $v_3$ between p+Au ($^{3}$He+Au) and d+Au collisions with a similar number of particles produced in the collisions ($\langle \mathrm{N_{ch}} \rangle$). 
The difference in the final state contributions, such as hydrodynamic evolution of the systems, are expected to be largely canceled in these ratios~\cite{Liu:2018xae}, 
making them the ideal probes of the initial geometry of the collisions. 
The $v_2$ and $v_3$ ratios are compared to the ratios of eccentricities ($\epsilon_2$ and $\epsilon_3$) of the initial energy density spatial distributions from Glauber model calculations. 
Model $a$ uses the default Glauber model to calculate the nucleon position and does not take into account the subnucleonic fluctuations~\cite{Nagle:2013lja}. 
Although the calculations from model $a$ can describe the measured $v_2$ ratio between $^{3}$He+Au and d+Au collisions, as well as the $v_3$ ratio between p+Au and d+Au collisions, 
they are $6.0\sigma$ ($6.2\sigma$) away from the $v_2$ ($v_3$) ratio between p+Au ($^{3}$He+Au) and d+Au collisions. 
On top of model $a$, model $c$ utilizes a Gaussian-like gluon field surrounding the valence quarks inside the nucleons in the nuclei~\cite{Welsh:2016siu} to include subnucleonic fluctuations. 
The calculations from model $c$ agree with the measured ratios of $v_2$ and $v_3$ in all the collision systems. 

These comparisons reveal significant influences from subnucleonic fluctuations in the initial geometry of p+Au, d+Au and $^{3}$He+Au collisions. 
The experimental measurements indicate that such fluctuations are crucial to understand the collective behavior in small collision systems, providing stringent constraints on the theoretical parameters regulating the subnucleonic fluctuations. 

Over the past several years, collisions of small systems have become a crucial probe of collective behavior and the conditions required for QGP formation. A central open question is whether the initial geometric and hydrodynamic conditions that drive QGP formation in large nuclei apply equally in small systems. In particular, the roles of subnucleonic fluctuations and nuclear deformation in shaping the initial geometry and seeding anisotropic flow must be understood to quantify the medium’s hydrodynamic response. Furthermore, resolving how these microscopic geometric features influence early-time dynamics and the approach to local thermalization in small systems will sharpen theoretical descriptions of QGP emergence. Upcoming high-precision measurements at RHIC and the LHC, combined with improved modeling of subnucleonic structure and fluctuating initial conditions, will allow stringent tests of the universality of QGP properties across system size and provide deeper insight into the mechanisms that drive the collective behavior.

\section {Electromagnetic probes and Ultra-peripheral collisions}

Electromagnetic probes, including real photons and dileptons, constitute the only direct penetrating probes of the hot and dense QCD medium produced in relativistic heavy-ion collisions~\cite{Shuryak:1978ij,McLerran:1984ay,Stoecker:1986ci,Rapp:2014hha,Kajantie:1981wg,Churchill:2023zkk}. Once emitted, they interact solely via the electromagnetic interaction and therefore propagate through the medium with negligible final-state interactions, retaining information on the space–time evolution and emission characteristics of the hot QCD matter. 
Ultra-peripheral collisions (UPCs) of heavy ions are characterized by impact parameters exceeding the sum of the nuclear radii, such that hadronic interactions are strongly suppressed~\cite{Bertulani:2005ru}. In this regime, the Lorentz-boosted electromagnetic fields of relativistic heavy nuclei can be treated as intense fluxes of quasi-real photons~\cite{vonWeizsacker:1934nji,Williams:1934ad}. These photons induce photon–photon and photon–nucleus interactions, providing a clean and well-controlled framework for precision studies of QED processes and nuclear parton distributions~\cite{Baltz:2007kq,STAR:2019wlg,STAR:2004bzo,PHENIX:2009xtn,ALICE:2013wjo,CMS:2018erd,CMS:2020skx,CMS:2023snh,CMS:2025oxg,ATLAS:2017fur,ATLAS:2019azn}.

\subsection{Probing QGP temperature via dileptons}
Among various experimental probes, dileptons (e.g. $\ee$ and $\uu$) pairs—have emerged as powerful messengers of the hot and dense QCD matter created in relativistic heavy ion collisions. 
The temperature of the emitting source governs the spectral hardness of the dilepton distribution: in general, higher temperatures produce harder energy and invariant-mass spectra~\cite{Rapp:2014hha}. In the QGP phase, temperatures can reach several hundred MeV—well above the QCD critical temperature $\Tc$~\cite{Gross:1980br,Linde:1978px,Roberts:2000aa}.
The critical temperature $\Tc$ marks the transition from hadronic matter to QGP. 
Lattice QCD calculations~\cite{HotQCD:2018pds,Borsanyi:2020fev} predict that, at finite baryon chemical potential ($\muB < 300$ MeV), this transition is a smooth crossover occurring at $\Tc \approx 156.5$ MeV. After the violent initial impact, the collision system undergoes rapid expansion and cooling, continuously emitting real photons and virtual photons (which subsequently materialize as dileptons) throughout its evolution.

While direct photons have long been used to estimate the QGP temperature at RHIC and the LHC~\cite{PHENIX:2014nkk,PHENIX:2022rsx,ALICE:2015xmh,PHENIX:2022qfp}, their spectra are subject to substantial Doppler effects due to collective expansion. This complicates the extraction of the true temperature of their thermal emission source, as relativistic flow modifies the observed photon's momentum spectra. In contrast, dileptons offer a unique advantage: their invariant mass $M_{ll}$ is Lorentz invariant and unaffected by Doppler shifts, providing a more direct thermometer of the hot QCD medium~\cite{Rapp:2014hha,Kajantie:1981wg,Churchill:2023zkk}.

At early times, when temperatures are highest, quark–antiquark annihilation dominates~\cite{Braaten:1990wp,Churchill:2023vpt}. 
As the medium cools toward confinement, partons hadronize, and dileptons originate primarily from the in-medium production and decay of short-lived $\rho^{0}$ vector mesons. Since the $\rho^{0}$ decays within $\sim$1.3 $fm/c$—far shorter than the typical fireball lifetime—its dilepton mass spectrum encodes information on in-medium hadronic modifications near the phase transition~\cite{Cassing:1999es,Rapp:1999ej,Rapp:2014hha}.

Over the past three decades, a broad experimental program has pursued thermal dilepton measurements over a wide range of beam energies, from a few GeV to several TeV~\cite{DLS:1989xpm, HADES:2019auv,CERESNA45:2002gnc,NA60:2007lzy,NA60:2008dcb,NA60:2009una,NA60:2016nad,STAR:2013pwb,STAR:2015zal,STAR:2015tnn,STAR:2023wta,PHENIX:2015vek,ALICE:2018ael}. The dilepton spectra measured in the low-mass region (LMR) have consistently revealed a broadened in-medium $\rho^{0}$ spectral function without a significant mass shift~\cite{Cassing:1999es,Rapp:1999ej,Rapp:2014hha}, signaling strong medium effects. At low beam energies, HADES experiment observed an exponential dielectron spectrum in Au+Au collisions at $\sqrt{s_{\rm NN}}=2.42$ GeV~\cite{HADES:2019auv}, suggesting near-thermal behavior with an effective temperature of $71.8\pm2.1$ MeV. In contrast, the NA60 experiment’s measurement of dimuons in In+In collisions at $\sqrt{s_{\rm NN}}=17.3$ GeV~\cite{NA60:2008dcb} extracted a temperature of $200\pm12$ MeV~\cite{Specht:2010xu}, significantly exceeding $\Tc$ and providing the first direct evidence of radiation from a deconfined QGP phase.
However, at top RHIC and LHC energies, thermal dielectron studies face major challenges from the large background of correlated semileptonic decays of open-charm hadrons, which dominate the intermediate-mass region (IMR). Consequently, QGP temperature extraction at these energies has remained elusive. 

Recent STAR measurements~\cite{STAR:2024bpc} take advantage of the second phase of the RHIC Beam Energy Scan (BES-II), which provides datasets an order of magnitude larger than those from BES-I~\cite{STAR:2015zal,STAR:2023wta} and accesses intermediate energies where physics background contributions from open charm are strongly suppressed. This analysis delivers the first determination of the QGP temperature at RHIC through the measurements of thermal dielectron ($\mathrm{e^{+}e^{-}}$) production in Au+Au collisions at $\sqrt{s_{\rm NN}} = 27$ and 54.4 GeV within a single experimental framework.

\begin{figure}
\centering
\includegraphics[width=0.49\textwidth]{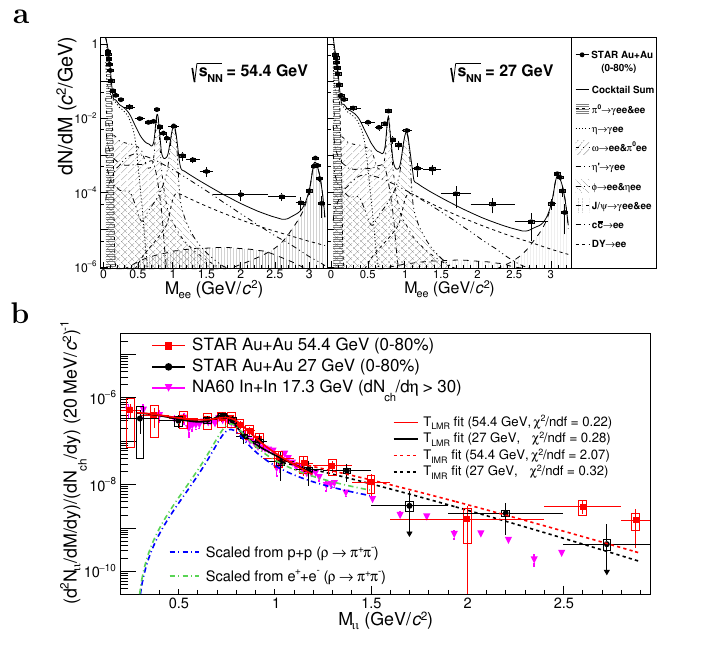}
\caption{Panel (a) shows the fully corrected inclusive dielectron spectra (black dots) and physics backgrounds (dashed and shaded lines) in Au+Au 0--80\% collisions at $\sqrt{s_{\rm NN}} = 54.4$ and $27$~GeV. Panel (b) shows the STAR thermal dielectron spectra compared with NA60 thermal dimuon spectra. Dashed curves show the temperature fits; dot-dashed curves indicate the expected vacuum $\rho$ from $p{+}p$~\cite{Aguilar-Benitez:1991hzq} and $e^{+}$+$e^{-}$~\cite{Derrick:1985jx} collisions. 
Vertical bars and boxes denote statistical and systematic uncertainties, and downward arrows indicate statistical uncertainties exceeding 100\%. The figure is taken from Ref.~\cite{STAR:2024bpc}.
}
\label{fig:dataVsCKT} 
\end{figure}

As shown in the top panel of Fig.~\ref{fig:dataVsCKT}, the inclusive dielectron spectra exhibit a clear enhancement above the total physics background (nonthermal sources), revealing a pronounced excess attributable to thermal radiation from the hot QCD medium. Subtracting the background yields the thermal dielectron spectra, presented in the bottom panel.
In the LMR, where thermal radiation is expected to be dominated by the in-medium $\rho^0$ decays, the excess spectra is fitted with a model combining an in-medium $\rho^{0}$ lineshape (relativistic Breit–Wigner) and a QGP continuum described by a Boltzmann distribution~\cite{Shuryak:2002kd,STAR:2003vqj,NA60:2016nad,NA60:2009una}. The extracted temperatures are $165 \pm 20\ \text{(stat)} \pm 21\ \text{(syst)}$ MeV at 27 GeV and $178 \pm 15\ \text{(stat)} \pm 13\ \text{(syst)}$ MeV at 54.4 GeV, consistent with the temperatures obtained from thermal dilepton measurements in STAR BES-I and NA60.
In the IMR, where partonic radiation dominates, the spectra follow a smooth QGP continuum described by the Boltzmann distribution $M^{3/2} e^{-M/k_B T}$ as suggested in~\cite{Rapp:2014hha,Churchill:2023zkk}. STAR obtains temperatures of $274 \pm 65 \text{ (stat)} \pm 10 \text{ (syst)}$ MeV at 27 GeV and $287 \pm 70 \text{ (stat)} \pm 34 \text{ (syst)}$ MeV at 54.4 GeV, consistent with NA60’s $245 \pm 17$ MeV result. The systematic difference between LMR and IMR temperatures reflects their origins: the IMR probes the earlier, hotter QGP phase, while the LMR is dominated by emission near the phase boundary.

\begin{figure}[h]
\centering
\includegraphics[width=0.49\textwidth]{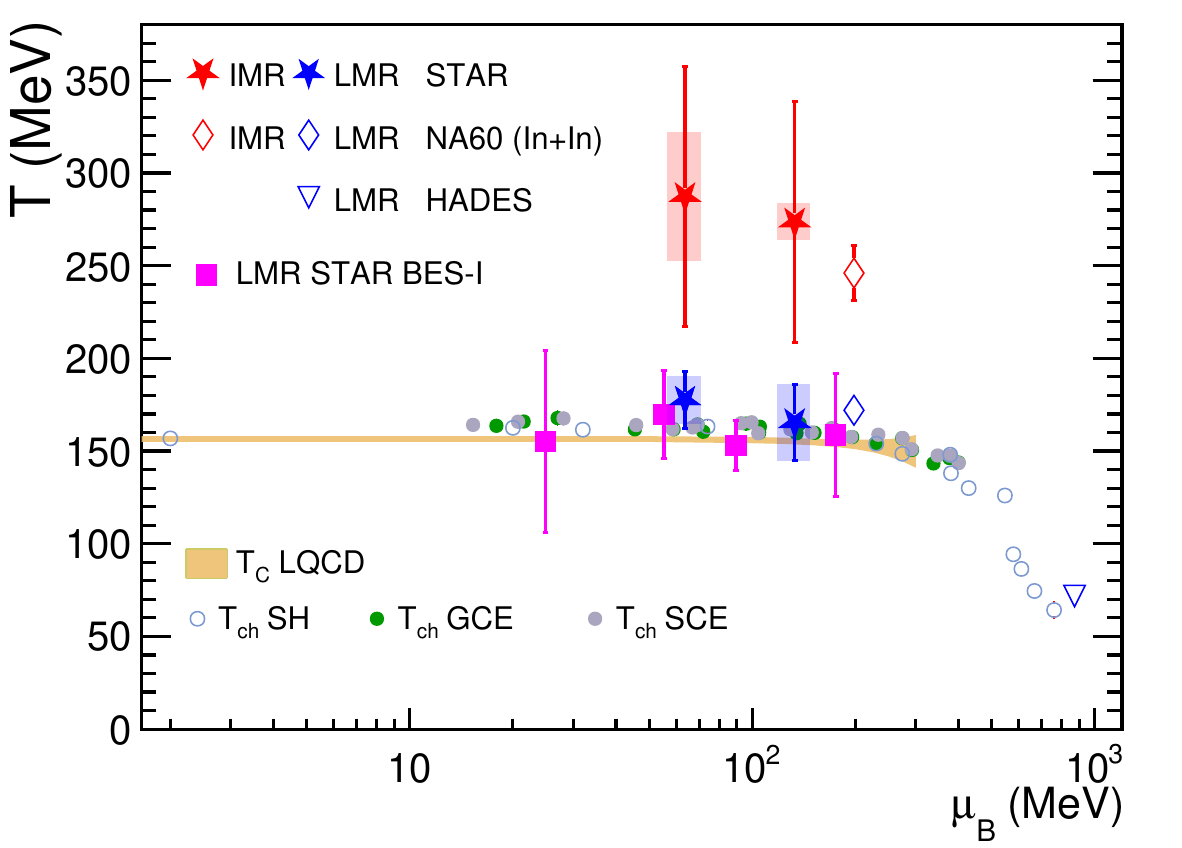}
\caption{Temperatures as a function of baryon chemical potential.
Temperatures measured by STAR are compared to those from NA60~\cite{NA60:2008dcb} (diamond markers) and HADES~\cite{HADES:2019auv} (inverted triangle markers). Chemical freeze-out temperatures from statistical thermal models (SH, GCE, SCE)~\cite{Andronic:2017pug,STAR:2017sal} (circles) and the lattice QCD critical temperature $\Tc$~\cite{HotQCD:2018pds} (yellow band) are also shown. All temperatures are plotted at their corresponding chemical freeze-out $\mu_B$. Vertical bars and boxes denote statistical and systematic uncertainties, respectively. The figure is taken from Ref.~\cite{STAR:2024bpc}.
}
\label{fig:TvsuB} 
\end{figure}

When plotted against the baryon chemical potential, the extracted dielectron temperatures reveal intriguing correlations. The LMR temperatures ($T_{\rm LMR}$) cluster around $\Tc$ and coincide with the chemical freeze-out temperatures ($T_{\rm ch}$) determined from hadron yields~\cite{Andronic:2017pug,STAR:2017sal}. This coincidence, long noted but not fully understood, suggests that the system spends considerable time radiating near $\Tc$, might due to the presence of a soft point in the QCD equation of state~\cite{Rischke:1996em,Nara:2021fuu}.

Thermal dileptons provide a unique and penetrating thermometer for the QGP created in high-energy nuclear collisions, despite their low production rates and substantial backgrounds. The new STAR measurements at 27 and 54.4 GeV represent the first multi-energy extraction of QGP temperatures at distinct evolutionary stages, yielding valuable constraints on the thermodynamics and phase structure of strongly interacting matter. 

Looking ahead, the large Au+Au data sets collected at multiple lower collision energies ($\sqrt{s_{\rm NN}} = 3.0-19.6$ GeV) during the STAR BES-II and the Fixed-Target program during 2019-2021, together with the $\sqrt{s_{\rm NN}}=200$ GeV sample (approximately $2\times10^{10}$ events) recorded in 2023–2025 with the upgraded inner TPC (iTPC)~\cite{YANG:2019rfi}, will enable substantially more precise investigations of critical phenomena near the QCD phase boundary using thermal dileptons. These data may also, for the first time, permit a determination of the QGP temperature in Au+Au collisions at RHIC’s top energy.

\subsection{Photon-photon and photon-nuclear interactions in UPCs}
During the past two decades, the STAR experiment has played a important role in developing UPC physics at RHIC~\cite{STAR:2002caw,STAR:2007elq,STAR:2017enh,STAR:2018ldd,STAR:2019yox,STAR:2021wwq,STAR:2022wfe,STAR:2024qpx}. Through measurements involving lepton-pair production, coherent and incoherent vector-meson photoproduction, quantum interference phenomena, and nuclear imaging, STAR has established UPCs as a unique experimental environment where strong-field QED, nuclear structure, and quantum entanglement can be studied. 

The Breit–Wheeler process, $\gamma + \gamma \rightarrow e^+ + e^-$, originally proposed in 1934, is the most fundamental mechanism for producing matter directly from light~\cite{Breit:1934zz}. In ultra-peripheral Au+Au collisions at $\sqrt{s_{\rm NN}}=200$ GeV, the Lorentz-boosted nuclei generate exceptionally intense photon fields, creating optimal conditions for observing the linear Breit–Wheeler process with quasi-real photons. Because two-photon dilepton production is a pure QED process, it serves as a clean benchmark for constraining photon flux models, validating the EPA, and probing possible higher-order QED effects relevant to strong electromagnetic fields~\cite{STAR:2019wlg,Li:2023yjt}.

STAR’s measurement, based on 6085 exclusive $e^+e^-$ pairs reconstructed in UPCs, provided the first experimental observation of the linear Breit–Wheeler process~\cite{STAR:2019wlg}. The measured distributions, such as invariant mass, rapidity, and transverse momentum, show excellent agreement with leading-order QED calculations using EPA photon fluxes, thereby tightening constraints on theoretical modeling. Following the initial confirmation at 200 GeV, STAR further extended the study of the Breit–Wheeler mechanism to lower energies during the Beam Energy Scan Phase II at $\sqrt{s_{\rm NN}}=54.4$ GeV~\cite{STAR:2024qpx}. This enabled a systematic comparison between experimental measurements and model predictions based on a Woods–Saxon nuclear charge distribution, ultimately yielding a novel method to constrain the charge distribution of heavy nuclei~\cite{Wang:2022ihj,STAR:2024qpx}. This application underscores the increasing role of UPC observables in advancing nuclear structure studies that are traditionally performed using low-energy electron scattering.

\begin{figure}[h]
\centering
\includegraphics[width=0.49\textwidth]{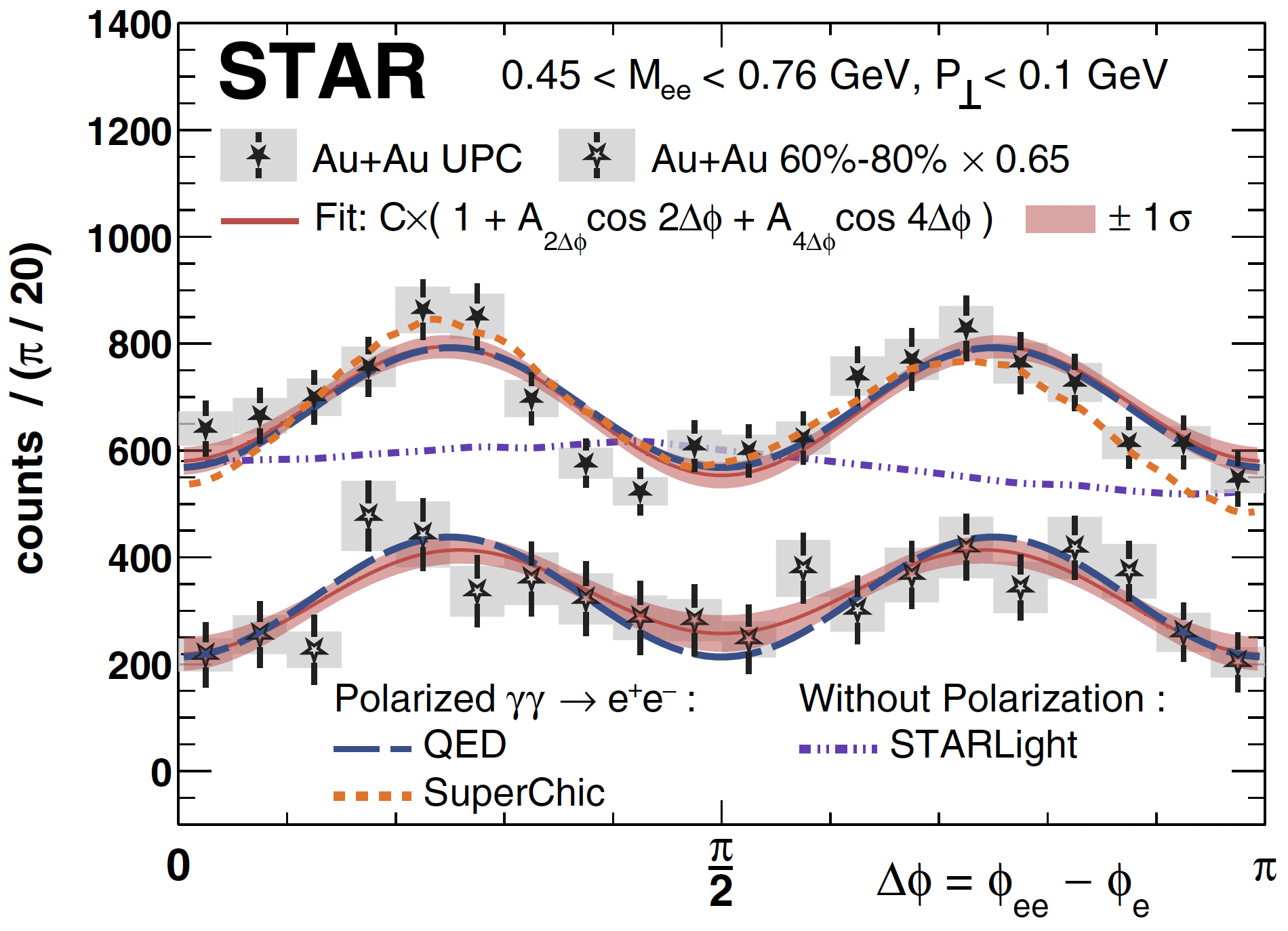}
\caption{Angular modulation of $\cos(4\Delta\phi)$ observed in Breit-Wheeler process at STAR. $\Delta\phi$ in defined as the angle between transverse momentum of $e^+e^-$ pair and single electron. Model calculations based on QED theory are shown as lines for comparison. Figure is from Ref.~\cite{STAR:2019wlg}.
}
\label{fig:4phi} 
\end{figure}

A particularly intriguing development concerns the polarization structure of the EM fields in UPCs. Due to Lorentz contraction, the quasi-real photons produced by relativistic nuclei are expected to carry linear polarization in the transverse plane. This will introduce a spin-induced orbital angular momentum (OAM) effect, from the polarization in $\gamma\gamma$ initial state to the angular modulations in the $e^+e^-$ final state~\cite{Li:2019sin,Li:2019yzy}. STAR observed a distinct fourth-order modulation, $\cos(4\Delta\phi)$ as shown in Fig.~\ref{fig:4phi}, where $\Delta\phi$ is defined as the angle between the transverse momentum of the pair and that of a single lepton~\cite{STAR:2019wlg}. The measurement represents experimental evidence of spin-induced OAM phenomena in photon–photon interactions, opening a promising avenue for studying polarization, angular momentum transfer, and quantum geometry in UPCs. Moreover, these findings highlight the broader potential of UPCs to explore polarization-dependent effects in both $\gamma\gamma$ and $\gamma A$ interactions.

$\gamma A$ processes have provided a complementary and highly sensitive tool for probing nuclear geometry and gluonic structure for decades~\cite{PHENIX:2009xtn,ALICE:2012yye,ALICE:2014eof,ALICE:2019tqa,ALICE:2021tyx,CMS:2018bbk,CMS:2022lbi,CMS:2023snh}. RHIC has measured coherent $\rho$-meson photoproduction in UPCs for almost twenty years; however, a persistent puzzle is that the nuclear radius of gold extracted from these measurements appears systematically larger than the accepted value~\cite{STAR:2002caw,STAR:2007elq,STAR:2017enh,PHENIX:2009xtn}. In effect, when imaged using high-energy photons, the nucleus seems “larger’’ than expected, raising questions about whether additional quantum or coherence effects modify the inferred nuclear profile~\cite{ma_new_2023}.

A major conceptual advance was made when STAR identified polarization dependent interference effects in coherent $\rho$ photoproduction. These results demonstrated unambiguously that the interpretation of $\gamma A$ measurements must incorporate both interference and photon polarization to avoid biased extractions of nuclear structure~\cite{STAR:2022wfe}. Since the impact parameter $b$ in UPCs is much larger than the nuclear radius, the photon’s linear polarization vector tends to align preferentially along the direction of $b$. As a result, in symmetric heavy-ion collisions, the two candidate photon emitters produce amplitudes that interfere in a $b$-dependent manner, modifying the observed production patterns~\cite{Zha:2020cst,Wu:2022exl}. By the spin interference effect, the initial spin direction can be probed by the transverse momentum of the vector meson, and the preferred alignment direction can be evaluated.

\begin{figure}[h]
\centering
\includegraphics[width=0.49\textwidth]{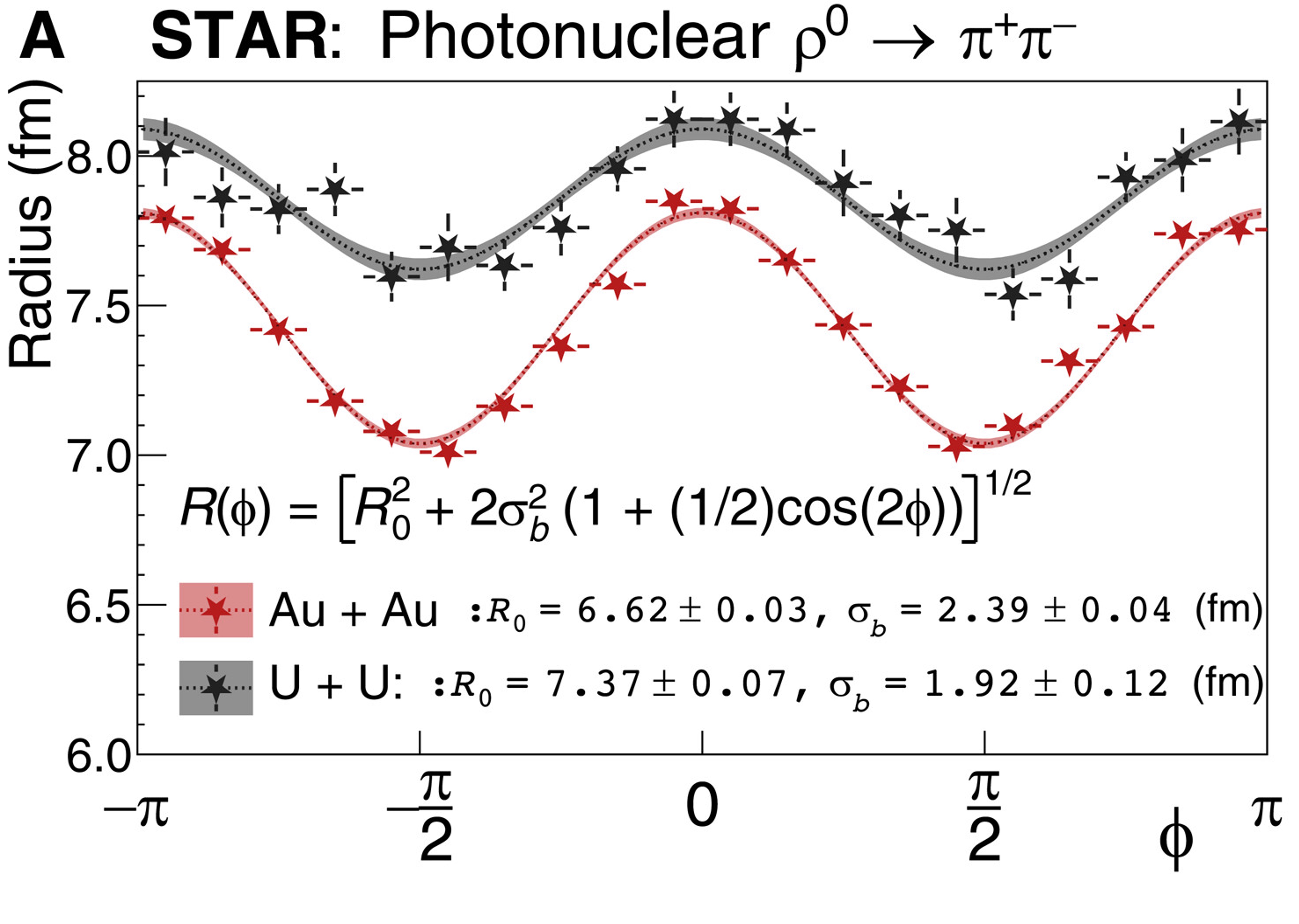}
\caption{Extracted mass radius as a function of polarization direction for gold and uranium. Radial parameter as a function of the polarization direction angle for Au + Au and U + U with an empirical second-order modulation fit. Figure is from Ref.~\cite{STAR:2022wfe}.
}
\label{fig:massradius} 
\end{figure}

By analyzing the transverse momentum of the $\rho$ meson together with the azimuthal correlation between the $\pi^+\pi^-$ decay products, STAR effectively selected different directions in the photon polarization space. As shown in Fig.~\ref{fig:massradius}, the extracted strong-interaction radius differs significantly across polarization orientations, with a maximal variation of nearly 0.9 fm. This demonstrates that the spin interference effect is not a small correction; rather, it is comparable in scale to the nuclear radius itself. Consequently, precision nuclear imaging with vector meson photoproduction requires careful accounting of both diffraction and polarization effects to avoid systematic biases.

The measurement can be interpreted as a femtoscopic realization of a double-slit interference experiment~\cite{ma_new_2023,Zha:2018jin}. In symmetric collisions ($A$+$A$), either nucleus may act as the photon source or as the scattering target, producing two amplitudes that interfere destructively at very low $p_T$, suppressing the yield near $t\approx 0$. Since the vector mesons are suggested to be produced on the nucleus, each nucleus itself acts as a slit on the Fermi scale. Remarkably, the separation between the two “slits’’ is orders of magnitude larger than the $\rho$ meson lifetime. The interference pattern is generated cooperatively by its decay products, $\pi^+\pi^-$. In contrast, in p+Au collisions, where the $\rho$ originates from a distinguishable photon emitter, the possibility of the photon coming from Au is $Z^2$ larger than that from protons, leading to the disappearance of the interference effect. Altering the charge of the projectile adjusts the trajectories of the photons, thereby inducing an effective observer effect; nevertheless, whenever interference patterns are detected, they still arise directly from coherent $\rho$-meson photoproduction. This cooperative pattern over space and time highlights the presence of quantum entanglement between production amplitudes~\cite{Cotler:2019gmg,Liu:2021hau,BESIII:2021ypr} , indicating that entanglement enables spin interference and offers a new method of 3D nuclear tomography.

Furthermore, the photon-induced physics program at RHIC can be extended well beyond polarization measurements. Using the high-statistics Au+Au data during the 2023 and 2025 RHIC runs, STAR can study dilepton production from photon–photon interactions as a function of collision centrality in hadronic events. The event–plane–dependent mean $p_{\rm{T}}$ of $e^+e^-$ pairs provides a novel probe of the electromagnetic properties of the QGP. In addition, systematic measurements of photoproduced vector mesons in UPCs can help to constrain nuclear gluon distributions at RHIC energies. By spanning a broad range of vector-meson masses, these studies access gluon evolution in nuclei as functions of $Q^2$ and the nuclear mass number $A$, further establishing RHIC as a unique laboratory for nuclear structure studies.

\section {Heavy flavor physics and jets }
Heavy flavor quarks and energetic partons are penetrating probes of the QGP. They are predominately produced in the initial hard scattering, interact strongly with the QGP when traverse it and hadronize in later stage before reach the detectors. The experimental studies of heavy flavor hadrons and jets production can provide critical information of the QGP. 
\subsection{Quarkonium sequential suppression}
Quarkonium, bound state of heavy quarks and its anti-quarks, can be dissociated and regenerated in the medium, thus their yields are sensitive to the properties of the QGP, making them ideal probes for studying the QGP characteristics. The suppression of ground state of charmonium ($J/\psi$) is extensively studied at RHIC energies; strong evidence of dissociation due to the static and/or dynamic screening of the potential of heavy quark pair is found. The suppression level of quarkonia in the QGP is believed to be related to their binding energies, with states of lower binding energy expected to be more easily suppressed. This phenomenon is commonly referred to as sequential suppression. The study of sequential suppression can shed new light on the production mechanism of quarkonium in heavy ion collisions and the extraction of QGP properties.

At the top RHIC energy, the STAR Collaboration has extensively study the production of charmonium and bottomonium in heavy ion collisions (see Ref.~\cite{Tang:2020ame} for a recent review). Recently, the STAR Collaboration made significant progress in the study of quarkonium sequential suppression in the QGP at RHIC energies thanks to the development of the STAR detectors and the improvement of event statistics. Sequential suppression of bottomonium has first been observed in Au+Au collisions~\cite{STAR:2022rpk}, and more recently, sequential suppression of charmonium has been observed in Ru+Ru and Zr+Zr collisions~\cite{STAR:2025imj}. These measurements demonstrate that in the hot and dense medium created in relativistic nucleus–nucleus collisions, quarkonium states with lower binding energies or larger radii are more susceptible to dissociation and therefore exhibit stronger suppression. The observation of the sequential suppression of bottomonium and charmonium at RHIC energy provides a bridge connecting the studies of heavy-flavor quarkonium sequential suppression at the lower SPS energy and the higher LHC energy, serving as a crucial step in understanding the energy dependence of QGP properties.

\begin{figure}
\centering
\includegraphics[width=0.48\textwidth]{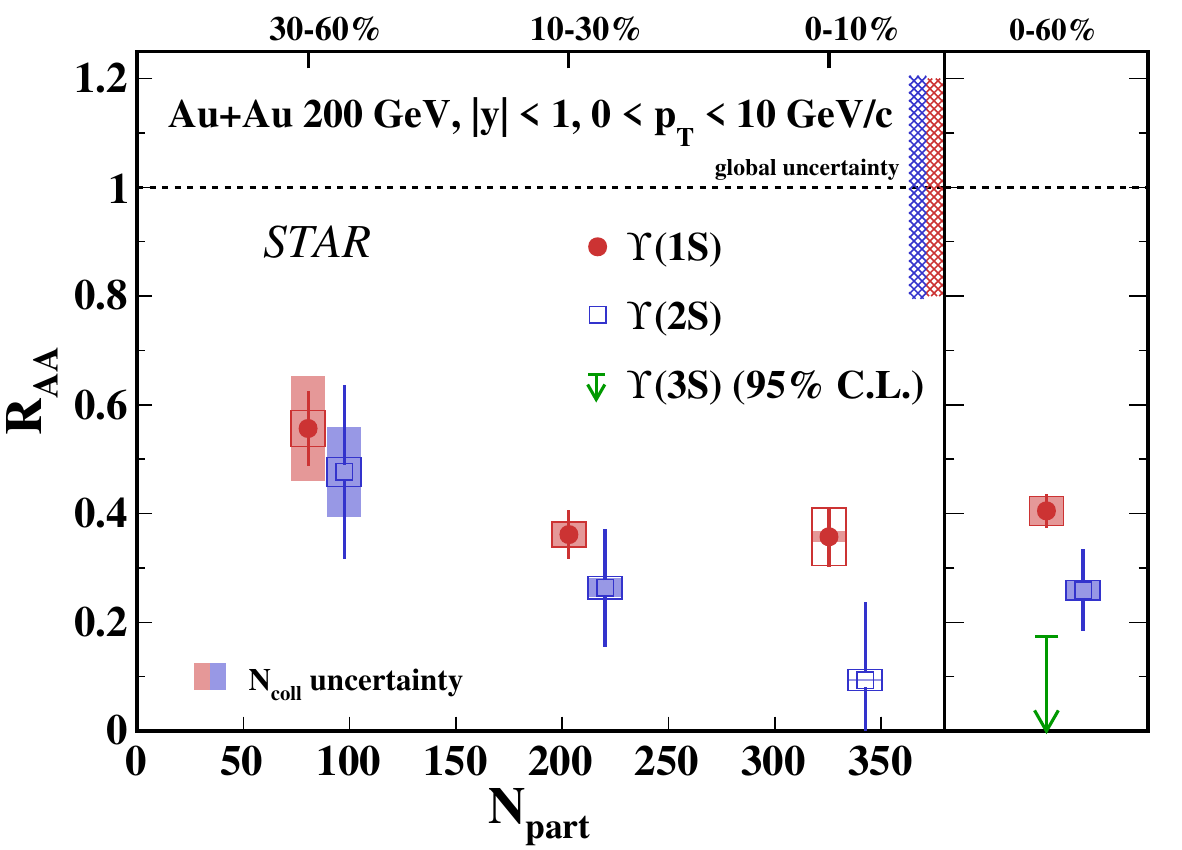}
\caption{Left: the $R_{\rm{AA}}$ of $\Upsilon$(1S) (circles) and $\Upsilon$(2S) (squares) as a function of $N_{\rm{part}}$ for $p_{\rm{T}} <$ 10 GeV/$c$. The statistical uncertainties are shown as the vertical bars, and the boxes indicate the systematic uncertainties. The Shadowed bands are the systematic uncertainties from $N_{\rm{coll}}$. The bands at unity stands for the global uncertainties. Right: the centrality integrated $R_{\rm{AA}}$ of various $\Upsilon$ states, including the 96\% upper limit for $\Upsilon$ (3S). Figure taken from Ref.~\cite{STAR:2022rpk}.}
\label{fig:upsilon}
\end{figure}

In the 0-60\% centrality Au+Au collisions at $\sqrt{s_{\rm{NN}}}=200$ GeV, the upper limit of the nuclear modification factor $R_{\rm{AA}}$ of $\Upsilon$ (3S) is estimated to be 0.17 with a 95\% confidence level (Fig.~\ref{fig:upsilon}), which is significantly more suppressed than $\Upsilon$ (1S). Even at the 99\% confidence level, the upper limit of the $\Upsilon$ (3S) $R_{\rm{AA}}$ is 0.26, still lower than the measured $R_{\rm{AA}}$ of $\Upsilon$ (1S) at the same centrality which is 0.4 $\pm$ 0.03 (stat.) $\pm$ 0.03 (sys.) $\pm$ 0.09 (norm.). The measured $R_{\rm{AA}}$ of $\Upsilon$ (2S) is 0.26 $\pm$ 0.08 (stat.) $\pm$ 0.02 (sys.) $\pm$ 0.06 (norm.), shows a hint that the suppression level of $\Upsilon$ (2S) is between $\Upsilon$ (1S) and $\Upsilon$ (3S). These results are consistent with the bottomonium sequential suppression pattern.

In addition to the sequential suppression of bottomonium observed in 200 GeV Au+Au collisions, the STAR Collaboration has observed the $\psi$(2S) signal in Ru+Ru and Zr+Zr collisions at the same energy for the first time. The $\langle N_{\rm part}\rangle$ dependence of inclusive $\psi$(2S) to J/$\psi$ double ratio shows a hint of decreasing trend from peripheral to central collisions, which is consistent with the expected stronger hot medium effect in central collisions. The results are consistent with the expected charmonium sequential suppression in heavy-ion collisions.

Measurements of bottomonium and charmonium states at $\sqrt{s_{\rm{NN}}}=200$ GeV show that states with lower binding energy are stronger suppressed in relativistic heavy-ion collisions, consistent with the picture of sequential suppression of quarkonia. These measurements fill the gap in sequential suppression studies at RHIC energies and provide crucial experimental data for investigating the energy dependence of QGP properties.


\subsection{Jets}
Investigating the underlying event characteristics linked to a high-$p_{\rm T}$ jet will enhance our understanding of QCD phenomena across both hard and soft energy scales at RHIC energies, and such measurements will offer valuable constraints to models of underlying event dynamics~\cite{Skands:2010ak}. 

STAR Collaboration has reported measurements of underlying event activity in $p + p$ collisions at $\sqrt{s}$ = 200 GeV at RHIC~\cite{STAR:2019cie}. The analysis followed the CDF topological structure method~\cite{CDF:2015txs}. For each collision event, the azimuthal angle ($\phi_{\mathrm{jet}}$) of the leading jet with the highest transverse momentum ($p_{\mathrm{T}}$) is designated as the reference angle. Reconstructed charged particles are subsequently grouped into distinct regions based on the difference between their azimuthal angle ($\phi_{i}$) and the reference angle, defined as $\Delta\phi = \phi_{i} - \phi_{\mathrm{jet}}$. Particles with $|\Delta\phi| < 60^\circ$ are assigned to the ``Toward'' region, while those with $|\Delta\phi - 180^\circ| < 60^\circ$ are categorized into the ``Away'' region, regardless of whether a jet is reconstructed in this area. The ``Transverse'' region encompasses particles with $60^\circ \leq |\Delta\phi| \leq 120^\circ$. The activity in each region is summed over the pseudorapidity range $|\eta| < 1$, which is inside of the TPC acceptance. The underlying event activity is evaluated by integrating the activity in the Transverse region over $\eta$ and is reported exclusively for charged tracks.

\begin{figure}[ht]
\centering
\includegraphics[width=0.48\textwidth]{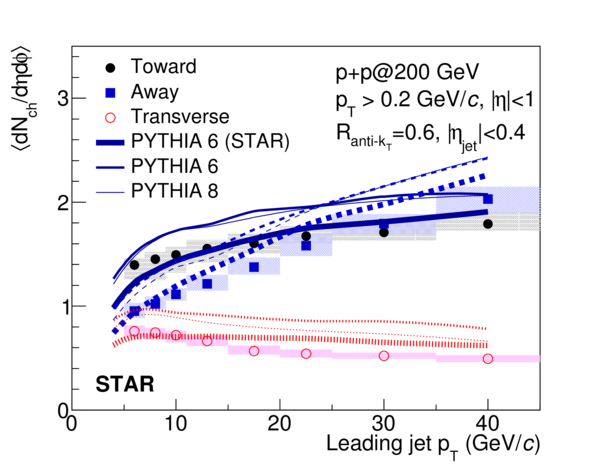}
\caption{The average charged particle multiplicity densities in the Toward, Away, and Transverse regions as functions of the leading jet $p_{\mathrm{T}}$, for charged particles with $p_{\mathrm{T}} > 0.2~\mathrm{GeV}/c$. The results are compared with three different PYTHIA simulations: the widest curves correspond to PYTHIA 6 (STAR), the intermediate-width curves represent the default PYTHIA 6 Perugia 2012 tune, and the thinnest curves depict the PYTHIA 8 Monash 2013 tune. The Toward region is represented by solid curves, the Away region by sparse dashed curves, and the Transverse region by dense dashed curves. Figure taken from Ref.~\cite{STAR:2019cie}.}
\label{fig_UE_multi}
\end{figure}

Figure~\ref{fig_UE_multi} shows the fully corrected average charged particle multiplicity densities for particles with $p_{\mathrm{T}} > 0.2~\mathrm{GeV}/c$ and $|\eta| < 1$ as a function of the leading jet $p_{\mathrm{T}}$. The statistical uncertainties are shown as vertical bars, which are smaller than the data symbols. The box heights represent systematic uncertainties, while their widths correspond to the $p_{\mathrm{T}}$ bin sizes. The average multiplicity densities in the Toward and Away regions, denoted by $\left\langle \frac{dN_{\mathrm{ch}}}{d\eta d\phi} \right\rangle$, exhibit an increasing trend with the leading jet $p_{\mathrm{T}}$. In contrast, for the Transverse region, $\left\langle \frac{dN_{\mathrm{ch}}}{d\eta d\phi} \right\rangle$ shows a slight decrease as the leading jet $p_{\mathrm{T}}$ increases. PYTHIA 6 (STAR), PYTHIA 6 default Perugia 2012, and PYTHIA 8 default Monash 2013 simulations are also shown as curves with widths from widest to thinnest, respectively. In the Transverse region, deviations between all simulations and the data are observed for jet $p_{\mathrm{T}} > 15~\mathrm{GeV}/c$. Among these, PYTHIA 6 (STAR), which was tuned using STAR's published minimum bias spectra, exhibits the closest agreement with the measured results.

\begin{figure}[ht]
\centering
\includegraphics[width=0.48\textwidth]{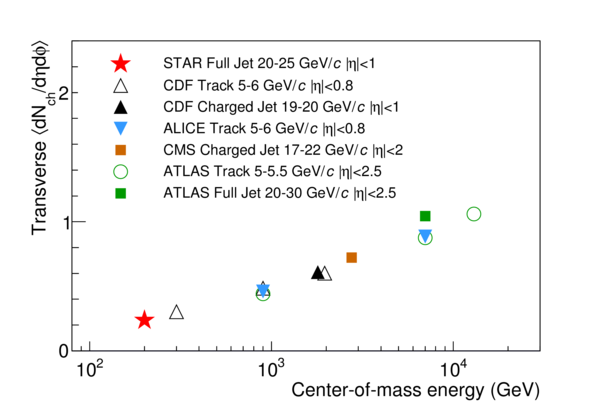}
\caption{Transverse charged particle densities at various collision energies. Figure taken from Ref.~\cite{STAR:2019cie}.}
\label{fig_UE_energy}
\end{figure}

In addition to the dependence of the underlying event activity on the energy scale of the hard scattering, investigating the collision energy dependence further enhances our understanding of low-momentum process modeling. Detailed measurements of the underlying event have been reported at various collision energies~\cite{DeRoeck:2005sqi}, enabling the study of the energy scaling of underlying event phenomena. Figure~\ref{fig_UE_energy} illustrates the dependence of the charged particle density on the collision energy for the Transverse regions. It is important to note that all reported results use charged particles with $p_{\mathrm{T}} > 0.5~\mathrm{GeV}/c$ to quantify the underlying event activity. The results reveals a near-linear increase with the logarithm of the collision energy, despite differences in the reference frameworks and pseudorapidity coverage across the measurements. This investigation into the collision energy dependence of the underlying event's average charged particle density offers additional insights that may help to correctly modeling the collision energy scaling of $p + p$ data. 

This study provides critical baseline measurements of the underlying event in $p+p$ collisions, which combined with previously reported minimum bias observables delivers valuable input and constraints for the predominantly phenomenological modeling of low-momentum QCD processes in Monte Carlo event generators. It also helps disentangle hard-scattering processes from soft background contributions, thereby reducing uncertainties in jet related measurements. These insights refine theoretical models to better simulate interactions between jets and the QGP medium~\cite{Aguilar:2021sfa, Yang:2022nei}, thereby enhancing the precision of jet tomography as a probe for QGP properties in heavy-ion collisions.

Hard probes, such as heavy flavor and jets, require exceptionally high statistics due to their low production cross sections. The STAR experiment’s Run 23 and 25 are projected to collect a total of 17 billion minimum bias events, offering an large and statistically powerful dataset. With such a vast dataset, measurements of hard probes will allow STAR to address essential questions about the inner workings of the QGP.  The resulting high-precision data will place stringent constraints on energy loss mechanisms and transport properties of the QGP, offering new insights into its internal dynamics.

\section {Antimatter Hypernucleus}

The properties of antimatter are of great interest since the matter-antimatter symmetry breaking in the early Universe is the precondition for the existence of the matter world we observe today. 
Antimatter is rare in the world surrounding us, because it can easily annihilate with the matter dominantly existing in the current Universe.
Antihypernucleus, as an antimatter partner of hypernucleus, is also difficult to produce and study in the laboratory.
Since the “negative energy solution” of the Dirac equation in 1928 predicted the existence of antimatter, scientists have only discovered eight antimatter (hyper)nuclei in nearly a century.

Relativistic heavy-ion collisions in STAR can generate the QGP with nearly equal quantities of quarks and antiquarks.
The collision system then expands, cools down, and transits to a hadron gas, producing (anti-)nucleons and (anti-)hyperons along with other particles.
Finally, (anti-)hyperons and (anti-)nucleons can coalesce to form (anti-)nuclei and (anti-)hypernuclei during the final phases of the evolution of the collision system~\cite{Chen:2010zzc}. 
Thus relativistic heavy-ion collisions are an efficient tool for producing and studying antinuclei and antihypernuclei~\cite{Chen:2018tnh,Chen:2024eaq}.

\begin{figure}[ht]
\centering
\includegraphics[width=0.48\textwidth]{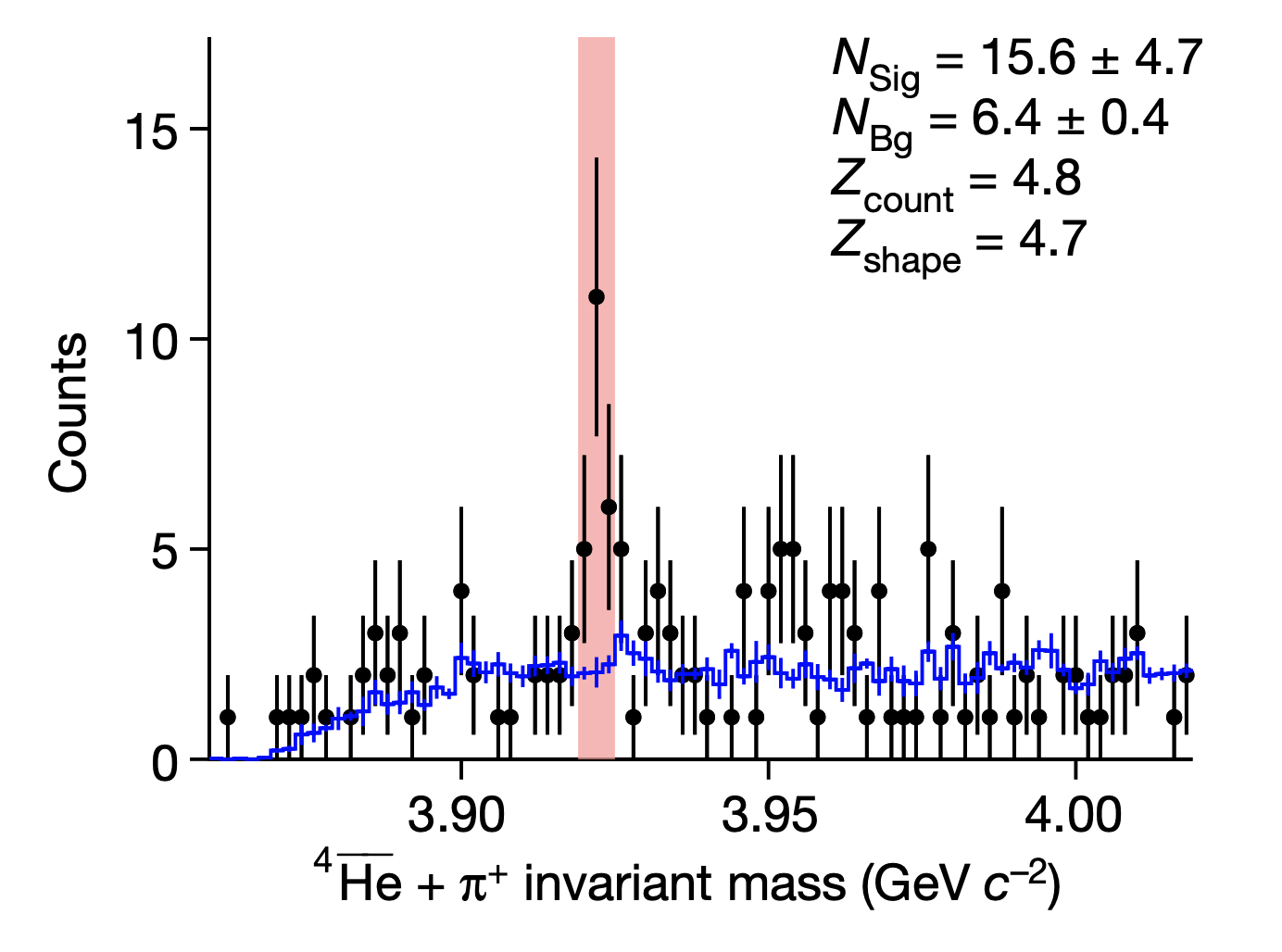}
\caption{Invariant mass distribution of $^{4}\overline{\hbox{He}}$ + $\pi^{+}$ with the \ahfl~signal peak in the expected signal region marked by the red band. The blue histogram shows the background distribution estimated by rotating one of the decay daughter particle. Figure adapted from Ref.~\cite{STAR:2023fbc}.}
\label{fig:antihyperH4}
\end{figure}

STAR Collaboration has discovered anti-hypertriton and anti-alpha in 2010 and 2011, 
respectively~\cite{STAR:2010gyg,STAR:2011eej}.
Recently, antihyperhydrogen-4 (\ahfl) has also been observed in the STAR experiment~\cite{STAR:2023fbc}.
A total of about 6 billion top RHIC energies collisions of Au+Au, U+U, Ru+Ru and Zr+Zr events recorded by STAR since 2010 are used in this search.
\ahfl~is reconstructed through its two-body decay into $^{4}\overline{\hbox{He}}$ and $\pi^{+}$.
Decay topological selection is applied, requiring the two daughters to come from a common decay vertex displaced from the collision point, in order to suppress the large amount of combinatorial background dominated by pairs of primary particles~\cite{STAR:2023fbc}. 15.6 \ahfl~signal candidates are observed on top of an estimated background count of 6.4, as shown in Fig.~\ref{fig:antihyperH4}. \ahfl~is the heaviest antimatter (hyper)nucleus observed to date. 

\begin{figure}[ht]
\centering
\includegraphics[width=0.48\textwidth]{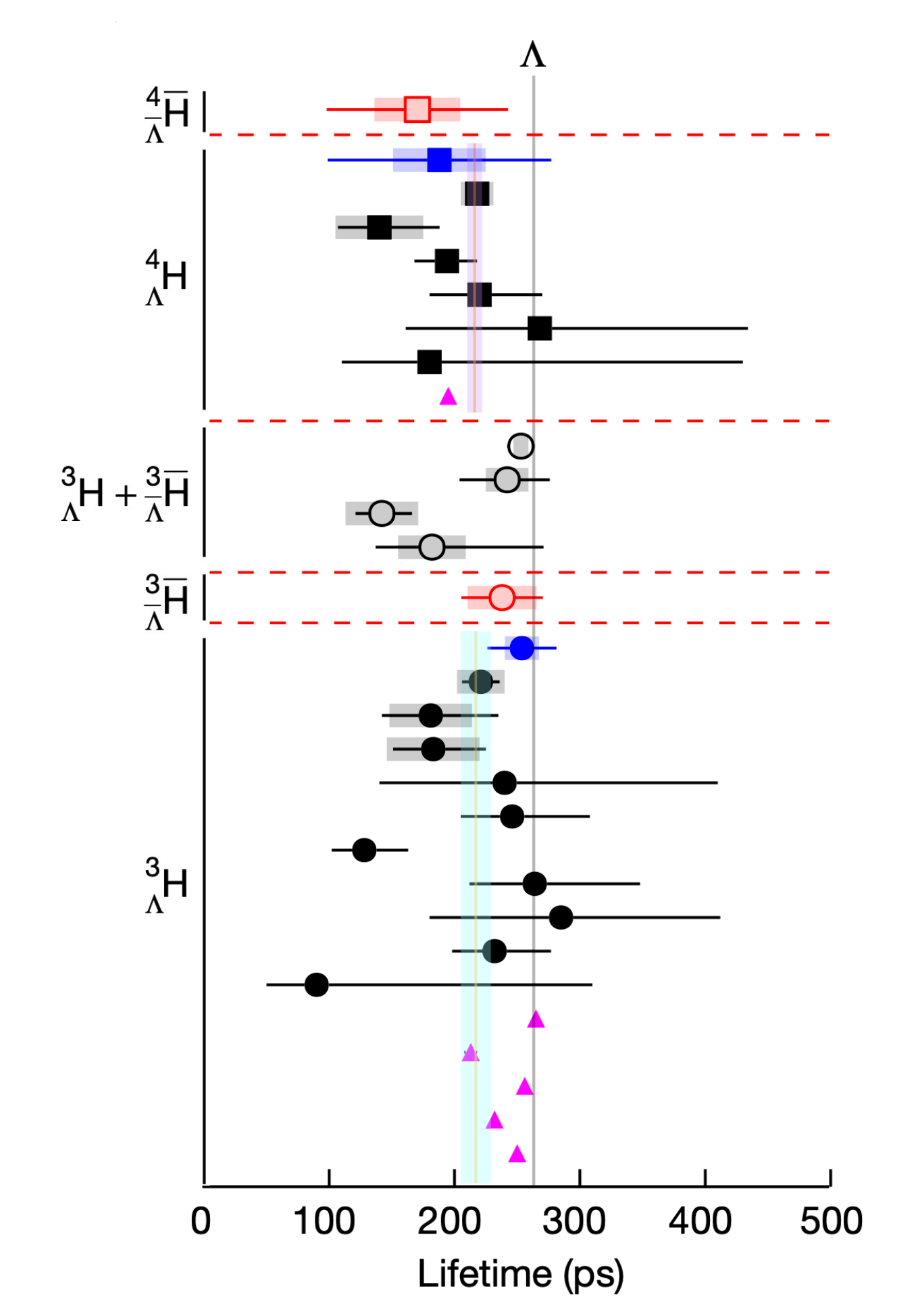}
\caption{Experimental measurements (circles and squares) and theoretical predictions (triangles) of the lifetimes of $^4_{\bar{\Lambda}}\overline{\hbox{H}}$, $^4_{\Lambda}\hbox{H}$, $^3_{\bar{\Lambda}}\overline{\hbox{H}}$ and $^3_{\Lambda}\hbox{H}$ from Ref.~\cite{STAR:2023fbc} and references therein.
Error bars and boxes show statistical and systematic uncertainties,
respectively.
Solid vertical lines with shaded regions represent the average lifetimes of $^3_{\Lambda}\hbox{H}$ and $^4_{\Lambda}\hbox{H}$ and their corresponding uncertainties.
The vertical gray line shows the lifetime of the free $\Lambda$.}
\label{fig:hypernucleusLifetime}
\end{figure}

To test the symmetry between the properties of the particles of matter and antimatter, the lifetimes of $^3_{\bar{\Lambda}}\overline{\hbox{H}}$ and \ahfl~ have been measured and compared to their corresponding matter counterparts, as shown by the red and blue markers in Fig.~\ref{fig:hypernucleusLifetime}. 
The lifetime differences between hypernuclei and their corresponding anti-hypernuclei are $\tau\left({^3_{\Lambda}\hbox{H}}\right)-\tau\left({^3_{\bar{\Lambda}}\overline{\hbox{H}}}\right)=$[16 $\pm$ 43(stat.) $\pm$ 20(sys.)] ps and ~$\tau\left({^4_{\Lambda}\hbox{H}}\right)-\tau\left({^4_{\bar{\Lambda}}\overline{\hbox{H}}}\right)=$[18 $\pm$ 115(stat.) $\pm$ 46(sys.)] ps. Both are consistent with zero, as expected from the CPT theorem. Although STAR did not observe any deviation from the expectations of the Standard Model, the detection of anti-nuclei has extended the antimatter frontier explored by humanity~\cite{Chen:2018tnh}.

\section {Nuclear structure}

\begin{figure*}[t]
\includegraphics[width=0.75\textwidth]{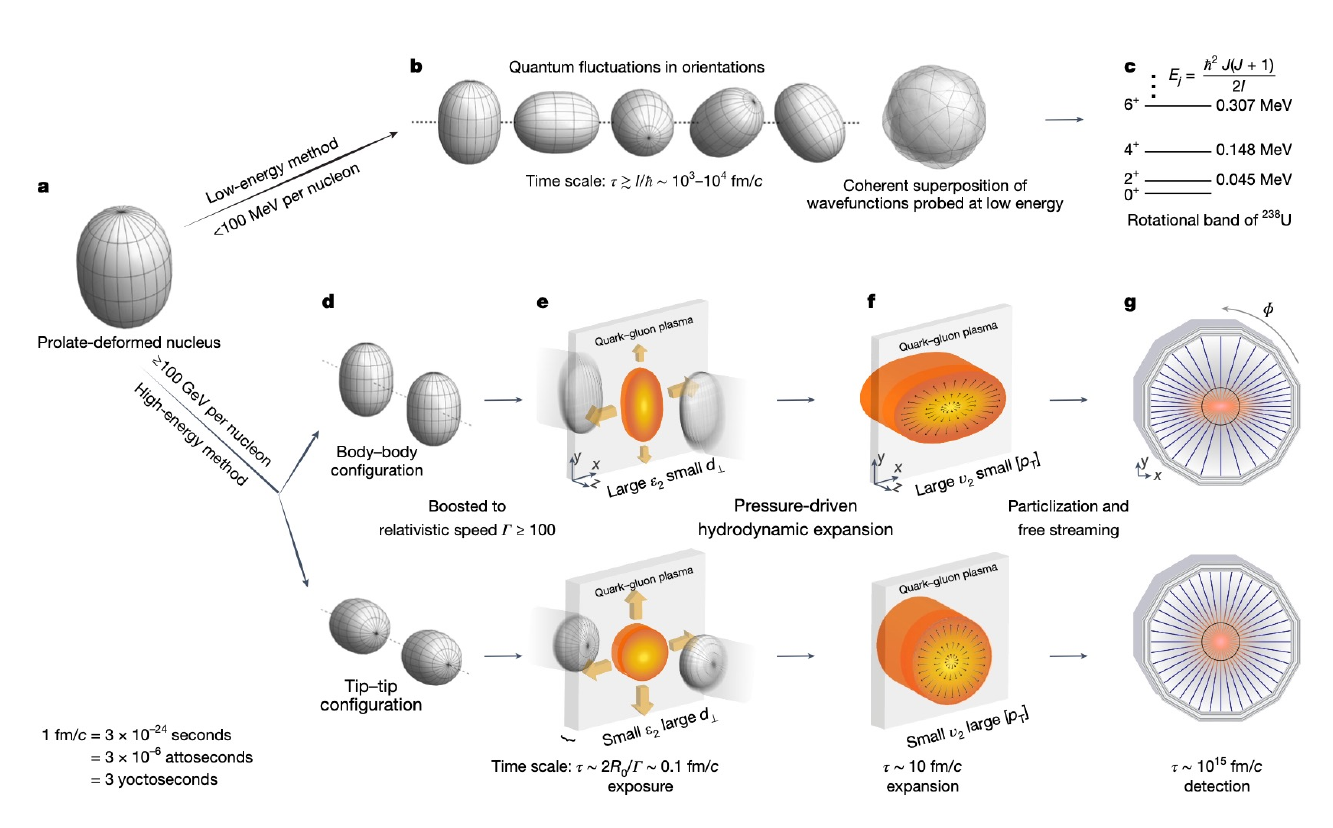}
\caption{Illustration of the methods for determining the nuclear shape in low and high energies. Cartoon of a well-deformed prolate-shaped nucleus (panel a). Quantum fluctuations over Euler angles for this nucleus and associated overall timescale (panel b). Quantum mechanical manifestation of the deformation in terms of the first rotational band of $^{238}$(panel c). Aligning the two nuclei in the head-on body-body configuration (top) and tip-tip configuration (bottom) (panel d). High-energy collision of two Lorentz-contracted nuclei and resulting 3D profile of the initially produced QGP, in which the arrows indicate the pressure gradients (panel e). The 3D profile of the QGP at the end of the hydrodynamic expansion before it freezes out into particles, in which the arrows indicate the velocities of fluid cells (panel f). Charged particle tracks measured in the detector (panel g). The timescales shown are in units of fm$/c$-the time for light to travel 1 femtometre. The body-body configuration has large eccentricity $\varepsilon_2$ and small gradient $d_{\perp}$, leading to large elliptic flow $v_2$ and smaller average transverse momentum $\left[p_{\rm{T}}\right]$ and vice versa for tip-tip configuration. Figure is taken from Ref.~\cite{STAR:2024wgy}.}
\label{fig:nature}
\end{figure*}

Atomic nuclei, the cores of visible matter, are self-organized quantum many-body systems whose shapes have long been a central focus of nuclear physics. Traditionally, nuclear shapes in the nuclear chart are probed using low-energy spectroscopic or scattering experimental techniques. Theoretically, for large enough nuclei, knowledge of a Woods–Saxon density function is to a great extent enough to capture the main properties of their structure and systematically explain the related experimental signatures. For light or intermediate-mass (A $\sim$ 20) nuclei, a mean-field description breaks down and two-body nucleon-nucleon correlations acquire prominent importance, as described by the first principle $ab$ $initio$ approaches. These methods operate on long timescales, where the instantaneous nuclear shape is obscured by long-timescale quantum fluctuations, making the direct observation of a nucleus’s instantaneous spatial matter distribution challenging. The STAR Collaboration introduced a novel, high-energy approach: the collective-flow-assisted nuclear shape-imaging method, which captures an instantaneous ``snapshot" of the nuclear geometry by analyzing the collective response of debris from ultra-relativistic heavy ion collisions, illustrated by Fig.~\ref{fig:nature}~\cite{STAR:2024wgy}.

This groundbreaking technique leverages the extreme conditions of a transient, hot, and dense state of matter, known as the QGP in ultra-relativistic heavy ion collisions. Due to extreme Lorentz contraction, the effective ``exposure time" of the collision is ultra-short ($\tau_{expo} \leq$ 0.1 fm/$c$), which is much faster than the characteristic rotational time of the nucleus. This ensures the capture of a collision-specific, instantaneous spatial distribution of nucleons. The initial shape of the colliding nuclei directly defines the initial geometry of the QGP. The QGP then undergoes rapid, pressure-gradient-driven hydrodynamic expansion, characterized by collective flow. This collective response translates the initial geometric anisotropy into momentum anisotropies of the final-state particles (the ``debris"), effectively providing an image of the initial nuclear shape~\cite{Giacalone:2021udy,Jia:2025wey}.

The experiment analyzed collisions of $^{238}\mathrm{U}+^{238}\mathrm{U}$ at $\sqrt{s_{\rm NN}} = $193 GeV and $^{197}\mathrm{Au}+^{197}\mathrm{Au}$ $\sqrt{s_{\rm NN}} = $200 GeV with similar sizes but distinct nuclear structures~\cite{STAR:2024wgy}. $^{238}\mathrm{U}$ serves as a benchmark due to its known highly prolate (elongated) shape, contrasting with the slightly oblate $^{197}\mathrm{Au}$. This shape difference enhances the sensitivity of the collective flow signal to deformation. The study focused on head-on collisions, where deformation effects are strongest, manifesting a crucial shape-size anti-correlation in $^{238}\mathrm{U}$ collisions. ``Body-body" configurations lead to an elongated QGP with large elliptic flow $\left(v_2\right)$ and less radial expansion, while ``tip-tip" configurations result in a compact, rounder QGP with small $v_2$ and more radial expansion. 

This anti-correlation is quantified by three key fluctuating observables: the mean squared elliptic flow $\left(\left\langle v_2^2\right\rangle\right)$, the mean squared fluctuation of the transverse momentum $\left(\left\langle\left(\delta p_T\right)^2\right\rangle\right)$, and the covariance between the two $\left(\left\langle v_2^2 \delta p_T\right\rangle\right)$. By calculating the ratio of these observables between $\mathrm{U}+\mathrm{U}$ and $\mathrm{Au}+\mathrm{Au}$ collisions $\left(R_{\mathcal{O}}\right)$, the complex QGP final-state effects are largely cancelled, linking the measurements more directly to the initial nuclear deformation parameters: the quadrupole parameter $\left(\beta_2\right)$ and the triaxiality parameter $(\gamma)$. By comparing the experimental data with state-of-the-art hydrodynamic models (IPGlasma+MUSIC and Trajectum), the study achieved a quantitative and simultaneous extraction of the ground-state deformation parameters for $^{238} \mathrm{U}$: $\beta_{\rm 2U}=0.286 \pm 0.025$ and $\gamma_U=8.7^{\circ} \pm 4.5^{\circ}$. The extracted $\beta_{2 U}$ value is in excellent agreement with the low-energy spectroscopic estimate ($\beta_{2 \rm U}=0.287 \pm 0.007$). This consistency confirms that the large-scale nuclear deformation is the dominant factor determining the nucleon distribution even on the ultra-short timescale of the collision. Moreover, analyzing these ratios as a function of nuclear shape also provides valuable insights into the energy deposition mechanisms.

Significantly, this work marks the first direct extraction of the average triaxiality parameter $\gamma_U$ of a nucleus in its ground state without relying on excited-state transitions. The small value of $\gamma_U$ indicates only a slight deviation from axial symmetry, ruling out the possibility of a large triaxial deformation for the uranium nucleus in its ground state.

\begin{figure}[t]
\includegraphics[width=0.475\textwidth]{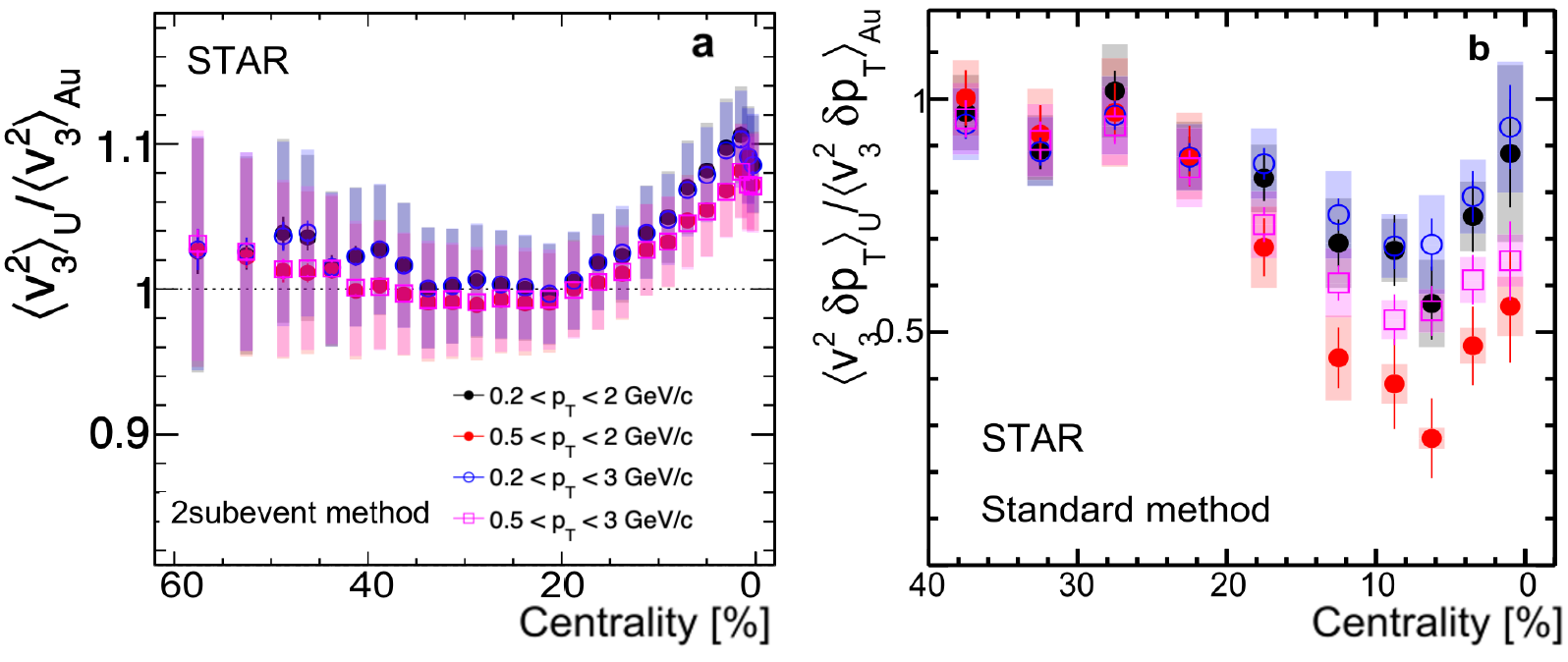}
\caption{Centrality dependence of ratios of $\left\langle v_3^2\right\rangle$($\mathbf{a}$) and $\left\langle v_3^2 \delta p_{\rm{T}}\right\rangle$($\mathbf{b}$) between U+U and Au+Au collisions in four $p_{\rm T}$ ranges. Figure is taken from Ref.~\cite{2025rot}.}
\label{fig:ropp}
\end{figure}
The direct measurement of higher-order deformations in the low-lying ground state of the nuclei presents a particularly crucial but challenging endeavor. The emergence of octupole deformation ($\beta_3$) in atomic nuclei leads to pear-like shapes, accompanied by spontaneous breaking of spatial reflection symmetry. This unique nuclear structure can significantly enhance the intrinsic electric dipole moment signal, which is closely linked to fundamental $\mathcal{CP}$ symmetry violation and may provide crucial insights into new physics beyond the Standard Model~\cite{Zhang:2025hvi,Zhang:2021kxj,Giacalone:2025vxa}. 

STAR Collaboration further extends their nuclear-shape-imaging method to specifically investigate the first experimental suggestion of a possible $\beta_3$ in the $^{238}$U. intrinsic $\beta_3$ of the colliding nucleus imprints a specific triangular anisotropy onto the initial geometry of the resulting QGP. This evidence is derived from observables involving triangular flow ($v_3$), which are largely insensitive to the dominant $\beta_2$ but are influenced by higher-order shape distortions~\cite{2025rot}. A surprising enhancement of $\left\langle v_3^2\right\rangle$ with $\sim 7 \%$ in central U+U collisions compared to Au+Au, contrary to simple size-scaling expectations. The ratio $R_{v_3^2}$ in four $p_{\rm T}$ ranges shows a slight increase in central collisions, possibly from in $\beta_{\rm 3,U}$, as shown in Fig.~\ref{fig:ropp}a. The ratio of $\left\langle v_n^2 \delta p_{\mathrm{T}}\right\rangle$ between U+U and Au+Au for standard method in four $p_{\rm T}$ intervals deviate significantly from unity in the ultra-central collision region in Fig.~\ref{fig:ropp}b. These deviations can be used to constrain the quadrupole and higher-order deformation of $^{238}$U. Hydrodynamic model comparisons (IP-Glasma+MUSIC) indicate that these patterns are consistent with an octupole deformation parameter of $\beta_{\rm 3,U} \sim 0.08-0.10$~\cite{Zhang:2025hvi}. Additionally, data also indicates comparable magnitudes between $\beta_{\rm 3,U}$ and $\beta_{\rm 4,U}$, implying the attractive potential to determine high-order deformations.

Note that the ``imaging-by-smashing" method as a potential discovery tool is particular essential to calibrate using the isobar collisions with the same nuclear mass number as an ideal case, $^{96}$Ru+$^{96}$Ru and $^{96}$Zr+$^{96}$Zr collisions~\cite{Giacalone:2025vxa,Jia:2022qgl,Zhang:2021kxj,Wang:2024mce}. STAR Collaboration also presented the first preliminary results for the observation of the $\beta_3$ and apparent large nuclear skin thickness ($a_0$) in the intermediate nuclei $^{96}$Zr (see details in Refs.~\cite{Xu:2022ikx,Jia:2022ozr}). Modern $ab$ $initio$ approaches to the nuclear many-body problem rooted in effective field theories of low-energy QCD are nowadays capable of addressing from first principles deformed intermediate-mass species and the emergence of clustering correlations therein~\cite{Zhang:2024vkh}. Recent STAR results demonstrated that accessing the possible clustering pattern in light nuclei $^{16}$O based on above ``imaging-by-smashing" method would be a new program of interdisciplinary QCD studies~\cite{STAR:2025ivi}. Determining the specific configurations of cluster patterns in light nuclei across energy scales is highly challenging, yet fundamentally important in nuclear physics.

In addition, understanding the three-dimensional structure of QGP remains one of the important open questions in heavy-ion physics. In this context, high-energy nuclear collisions, at the intersection of nuclear structure provide systematical, universal, and unbiased insight for probing this QGP’s 3D initial conditions~\cite{Zhang:2024bcb}. Immediate experimental applications at RHIC and LHC with various ion species, with extended detector acceptance, are expected to make these measurements feasible.

The flow-assisted nuclear shape imaging is a promising tool for exploring the structure of atomic nuclei in their ground state, applicable to any collision species~\cite{Jia:2022ozr}. These findings not only provide a more precise constraint on the initial conditions of QGP but also serve as a powerful new tool for investigating other essential questions in nuclear structure. Many applications are possible, such as nuclear shapes in odd-mass nuclides, hexadecapole deformations, dynamic deformations in soft nuclei, constraints on nuclear matrix elements in $0 \nu \beta \beta$ decay. This is also particularly important in small collision systems, where uncertainties in the properties and initial conditions of the QGP could reduce the sensitivity to nuclear structure differences between isobar or isobar-like species. It would be remiss not to mention that our approach promotes a critical convergence between high-energy nuclear physics and low-energy nuclear structure, expanding the scope of nuclear research across energy scales.

\section {Spin physics in p+p and A+A collision }

Understanding the nucleon spin structure remains a central frontier in QCD. Early deep inelastic scattering (DIS) experiments revealed that quark spins account for only a small fraction of the proton spin~\cite{EuropeanMuon:1987isl}, giving rise to the ``proton spin crisis.'' Subsequent theoretical and experimental efforts have focused on the contributions from gluon polarization, sea quark polarization, and orbital angular momentum~\cite{Ji:2020ena}.  

RHIC is currently the only polarized proton-proton collider in the world, and the STAR experiment has played a leading role in mapping out the nucleon spin structure~\cite{rhic_coldqcd_2023}. STAR benefits from large acceptance tracking, calorimetry, and forward detectors, enabling comprehensive studies of spin asymmetries and polarization observables. 

Recently, the global polarization effect in heavy ion collision has also become an important direction of high energy nuclear physics, as a new avenue to study the property of QGP. 
In this section, we focus on nucleon helicity distributions, transverse momentum dependent distributions, spin-dependent fragmentation, and spin physics in heavy ion collisions.  

\subsection{Nucleon helicity structure}

A major goal of the RHIC spin program is to determine the gluon spin contribution to proton spin, namely, the helicity distribution $\Delta g(x)$~\cite{Bunce:2000uv}. 
STAR has measured longitudinal double-spin asymmetries $A_{LL}$ for inclusive jet and di-jet production at $\sqrt{s}=200$ and 510~GeV~\cite{prl_jet_all_2015, prd_dijet_all_2017, prd_dijet_endcap_all_2018, prd_jet_all_510gev_2019, prd_jet_all_200gev_2021, prd_jet_all_500gev_2022, dijet_endcap_all_2025}. These measurements probe gluon momentum fractions in the range $x \sim 0.02$--0.3.  When included in global QCD analyses such as DSSV and NNPDF~\cite{rhic_coldqcd_2023, nnpdfpol2.0_jhep_2025}, these data confirm gluon spin $\Delta g(x)$ contribute significantly ($\sim$ 40\%) to proton spin, for example, 
the total contribution of $\Delta g(x)$ given by DSSV group is $\Delta G$ =0.22$_{-0.06}^{+0.03}$ for $x>0.05$ ~\cite{rhic_coldqcd_2023}. 

\begin{figure}[ht]
\centering
\includegraphics[width=0.45\textwidth]{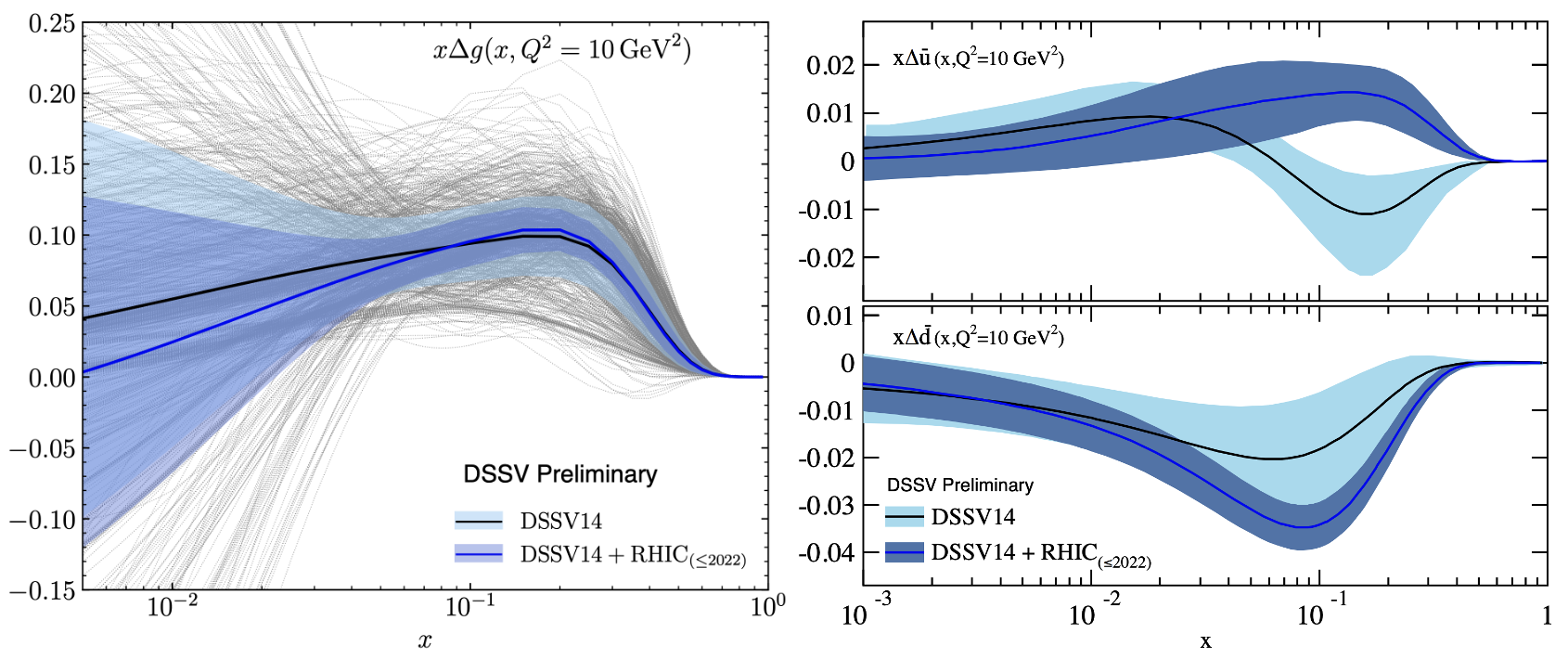}
\caption{The impact of the recent jet, pion and $W$ data on helicity distributions of gluon (left panel), anti-up and anti-down quark (right panel) at Q$^2$ = 10 GeV$^2$ from DSSV group. 
Figure taken from~\cite{rhic_coldqcd_2023}}
\label{fig:gluon}
\end{figure}

The flavor separation of nucleon spin structure, in particular the sea quark polarization, is one of the two initial motivations for the spin-physics program at RHIC~\cite{Bunce:2000uv}.
The helicity distributions of antiquarks can be probed with $W^{\pm}$ boson production in longitudinally polarized $pp$ collisions, without involving fragmentation function. 
STAR measurements of the single-spin asymmetry $A_L$ in $W$ production at $\sqrt{s}=500$~GeV allow unique flavor separation of $\Delta u$, $\Delta d$, $\Delta \bar{u}$, and $\Delta\bar{d}$~\cite{prl_w_al_2014, prd_w_al_2019,XU:2019iqd}. 
The sea quark $\Delta \bar{u}$ is now known to be positive and $\Delta\bar{d}$ is negative, as shown in right panel of Fig.~\ref{fig:gluon}. 
For the first time, those precision data from STAR show that there is a clear flavor asymmetry between $\Delta \bar{u}$, and $\Delta \bar{d}$~\cite{rhic_coldqcd_2023}. 

\subsection{Transverse spin asymmetry}

RHIC has made substantial breakthroughs in exploring transverse single spin asymmetries (TSSA) and transverse-momentum-dependent (TMD) distributions in polarized proton–proton collisions~\cite{Xu:2015bid,rhic_coldqcd_2023}.  
Understanding TSSA and TMD physics is a rapidly evolving field of QCD in recent years. 
Central to this program at STAR is the measurements of TSSA for hadrons, jets, and $W/Z$ boson with transversely polarized proton-proton collisions, in exploring TMD mechanisms including Sivers and Collins effect and high twist mechanism.
One major advance of initial-state effect is the measurement of the so-called Sivers effect, in which the intrinsic transverse momentum of a quark correlates with the transverse spin of the parent proton. Sivers effect has been studies with azimuthal asymmetries of weak boson~\cite{prl_w_an_2016,z_tssa_2023}, hadron and jets~\cite{prd_forward_an_2021,dijet_tssa_sivers_2023} productions in transversely polarized proton-proton collisions at STAR. 
These results provide an essential test of the predicted process-dependence (sign-change) of the Sivers function and the evolution of TMDs. 

\begin{figure}[ht]
\centering
\includegraphics[width=0.45\textwidth]{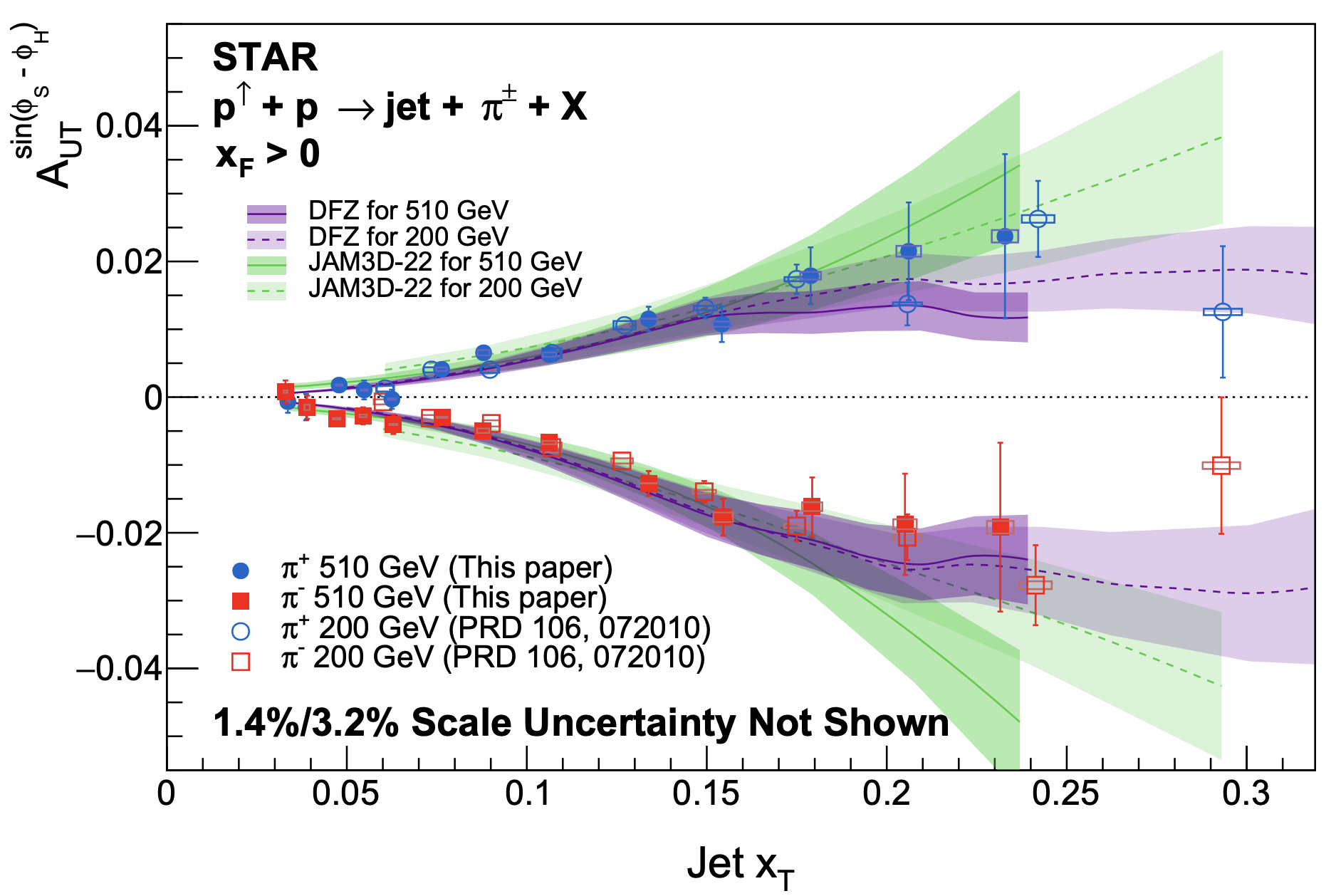}
\caption{Collins asymmetries, $A^{\mathrm{sin}(\phi_S - \phi_H)}_{UT}$, as a function of jet $x_T$ ($\equiv 2p_{T,jet}/\sqrt{s}$) for $\pi^{\pm}$ in p+p collisions at $\sqrt{s}$ = 510 GeV (solid points), compared with previous results at $\sqrt{s}$ = 200 GeV (open points) and theoretical calculations. 
Figure taken from~\cite{STAR:2025xyp}}. 
\label{fig:Collins}
\end{figure}

In particular, the azimuthal modulations or TSSA of hadron in jets in pp collision, are sensitive to the spin and transverse momentum correlation in the fragmentation process of a transversely polarized quark, which is known as Collins effect. 
STAR has measured Collins asymmetries for charged hadrons in jets in transversely polarized proton-proton collisions at $\sqrt{s}$ = 200 GeV and 500 GeV~\cite{prd_collins_500gev_2018,prd_collins_200_2022,STAR:2025xyp}. These data enable extraction of quark transversity distributions coupled with Collins fragmentation functions. 
Interestingly, as seen from Fig.~\ref{fig:Collins}, recent new results show a remarkable consistency for Collins asymmetries of $\pi^{\pm}$ in jets between $\sqrt{s}$ = 200 GeV and 510 GeV, when plotted versus energy-scaled distribution of jet transverse momentum, $x_{\mathrm{T}} = 2p_{\mathrm{T,jet}}/\sqrt s$. 
The asymmetry increases with jet $x_{\mathrm{T}}$, consistent with the expectation that magnitude of quark transversity grows with parton momentum fraction $x$.
As the asymmetries originate from the convolution of quark transversity with the Collins function, the comparison of Collins asymmetries from two energies at a fixed $x_{\mathrm{T}}$ (corresponding to a similar $x$) reflects the effect of TMD evolution.
This indicates that the Collins asymmetries are nearly energy independent with, at most, a very weak scale dependence in $pp$ collisions.
These results extend to high-momentum scales ($Q^2 \leq 3400$ GeV$^2$) and enable unique tests of evolution and universality in the TMD formalism.


\subsection{{$\Lambda$} polarization in pp collision}

Recent measurements at STAR have significantly advanced our understanding of hyperon spin phenomena in polarized and unpolarized proton-proton collisions. In a comprehensive study of spin transfer at $\sqrt{s}=200$~GeV, STAR reported measurements of longitudinal and transverse spin transfer coefficients, $D_{LL}$ and $D_{TT}$, for $\Lambda$ and $\bar{\Lambda}$ hyperons in longitudinally and transversely polarized pp collisions~\cite{prd_lambda_dll_2018,prd_lambda_dtt_2018,spin_transfer_2024}. These results provide important constraints on the polarized strange-quark distributions and the polarized fragmentation functions, also baselines for future measurements of hyperon polarization in lepton-nucleon scattering at EIC and EicC~\cite{Kang:2021kpt,Chen:2021zrr,Ji:2023cdh}. 

Recently, STAR report the first measurement of $\Lambda$ and $\bar{\Lambda}$ transverse polarization inside jets in unpolarized proton-proton collisions at $\sqrt s=200$ GeV~\cite{prl_lambda_in_jet_2025}.
The measured polarization relative to jet axis actually probes the polarizing fragmentation function (PFF), which describes the production of transversely polarized hadrons
from the fragmentation of an unpolarized parton~\cite{Mulders:1995dh}. 
PFF could be the main source of the surprisingly large transverse polarization of $\Lambda$ hyperons relative to the production plane in unpolarized hadron-hadron collisions observed more than 40 years ago~\cite{Bunce:1976yb,Liang:1997rt}.
A clear jet $p_T$
dependence is observed for $\Lambda$ polarization in jets.
This measurement provides the first experimental constraint on gluon-initiated polarizing fragmentation function, and opens a new window to study spin-dependent hadronization in jet environments.

Most recently, STAR reported the first measurement of spin correlations between $\Lambda$ and $\bar{\Lambda}$ pairs produced in $p+p$ collisions~\cite{nature_spin_correlation_2025}. A significant spin-correlation signal,  about 18\% for closely aligned pairs, was observed and interpreted as evidence for spin correlation of $s\bar{s}$ quark pairs surviving through hadronization.

\subsection{ Global polarization in heavy-ion collisions}

The landmark observation by the STAR Collaboration of nonzero global $\Lambda$/$\bar{\Lambda}$ polarization~\cite{the_star_collaboration_global_2017} in heavy-ion collisions has established spin measurement as an essential tool of the QGP study, as first proposed by Liang and Wang 20 years ago~\cite{Liang:2004ph,Liang:2004xn}.
More measurements have been performed for different hyperons and at varied collision energies, and the measured $\Lambda$ polarization shows a clear increase toward lower collision energies, consistent with the expectation from thermal vorticity in hydrodynamic models~\cite{Xu:2023lnd,Chen:2024aom}. 

\begin{figure}[htb]
    \centering
    \includegraphics[width=0.8\linewidth]{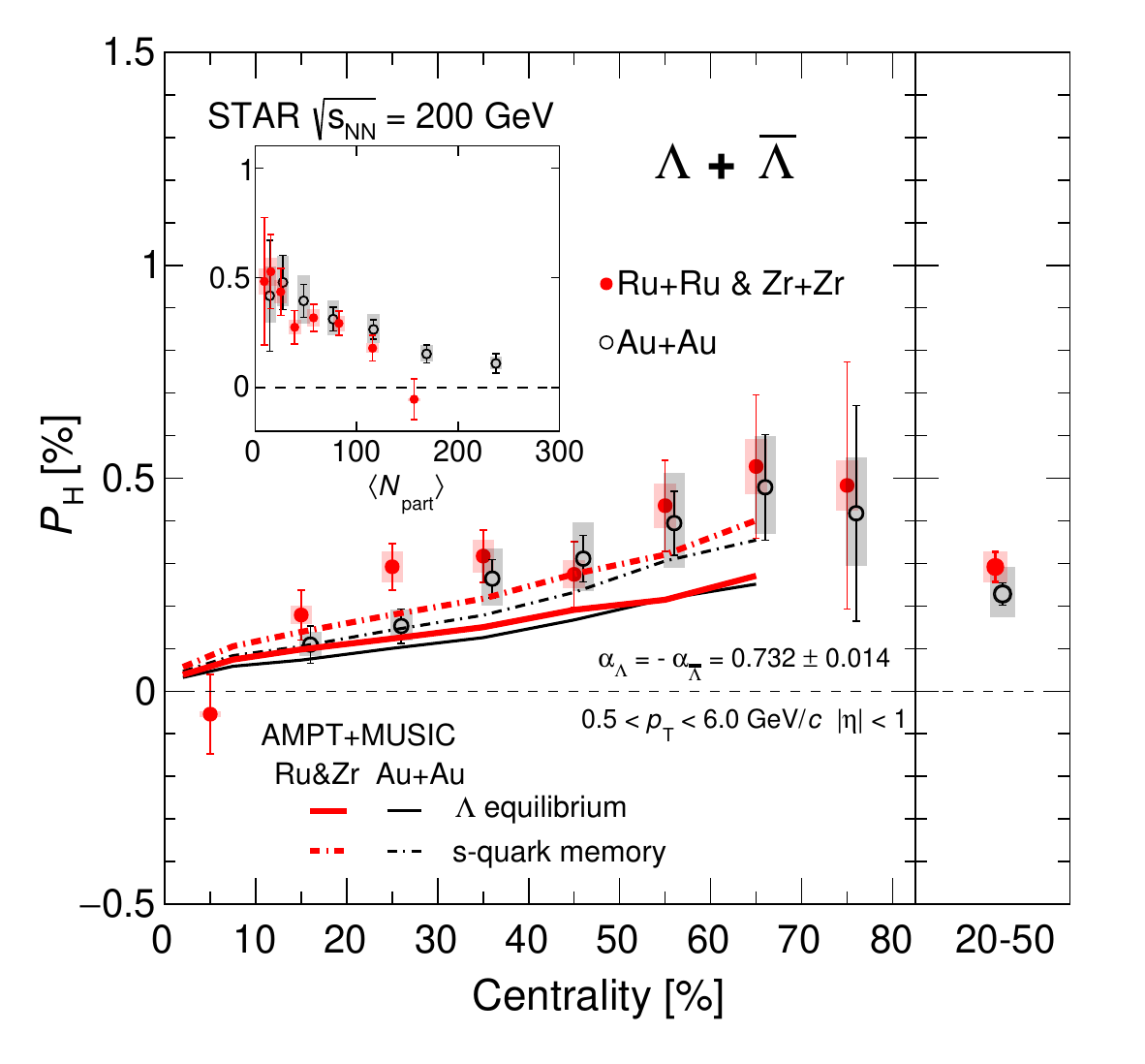}
    \vspace{-3mm}
    \caption{ 
    Global polarization of $\Lambda$+$\bar\Lambda$ as a function of centrality in Ru+Ru and Zr+Zr collisions at $\sqrt s$ = 200 GeV (left panel) and results in 20-50\% centrality (right panel), compared with results in Au+Au collisions. 
    The inset presents the same data plotted as a function of the average number of participants $\langle N_{\rm part} \rangle$. 
    Solid and dashed-dotted lines show calculations from the hydrodynamic model (MUSIC).
    Figure taken from \cite{global_isobar_2025}.
    }
    \label{fig:GlobalP_SystemSize}
\end{figure}

Recently, the global polarization of $\Lambda$, $\bar{\Lambda}$, $\Xi^-$, and $\bar{\Xi}^+$ hyperons has been measured in isobar collisions of Ru+Ru and Zr+Zr at $\sqrt{s_{\rm NN}}$=200 GeV at STAR~\cite{global_isobar_2025}. 
Hydrodynamic calculations predict a larger polarization in smaller collision systems due to a shorter fireball lifetime. 
As shown in Fig.~\ref{fig:GlobalP_SystemSize}, the $\Lambda$+$\bar\Lambda$ global polarization in isobars are comparable to the results in Au+Au collisions at the same energy, with a light hint of larger polarization in isobar collisions. 
The global polarization was also measured in Ru+Ru and Zr+Zr systems separately and found to be consistent with each other.
Also, the measurements show no significant difference between $\Lambda$ and $\bar\Lambda$ in isobar collisions, i.e, no splitting due to potential effect from the magnetic field is observed~\cite{global_isobar_2025}.
Furthermore, measurements of multi-strange hyperons ($\Xi$ and $\Omega$) in isobar collisions indicate a possible spin hierarchy ($P_{\Omega} > P_{\Xi} > P_{\Lambda}$), though higher-statistics are required for a solid conclusion~\cite{global_isobar_2025}.

\begin{figure}[thb]
\begin{center}
\includegraphics[width=0.85\linewidth]{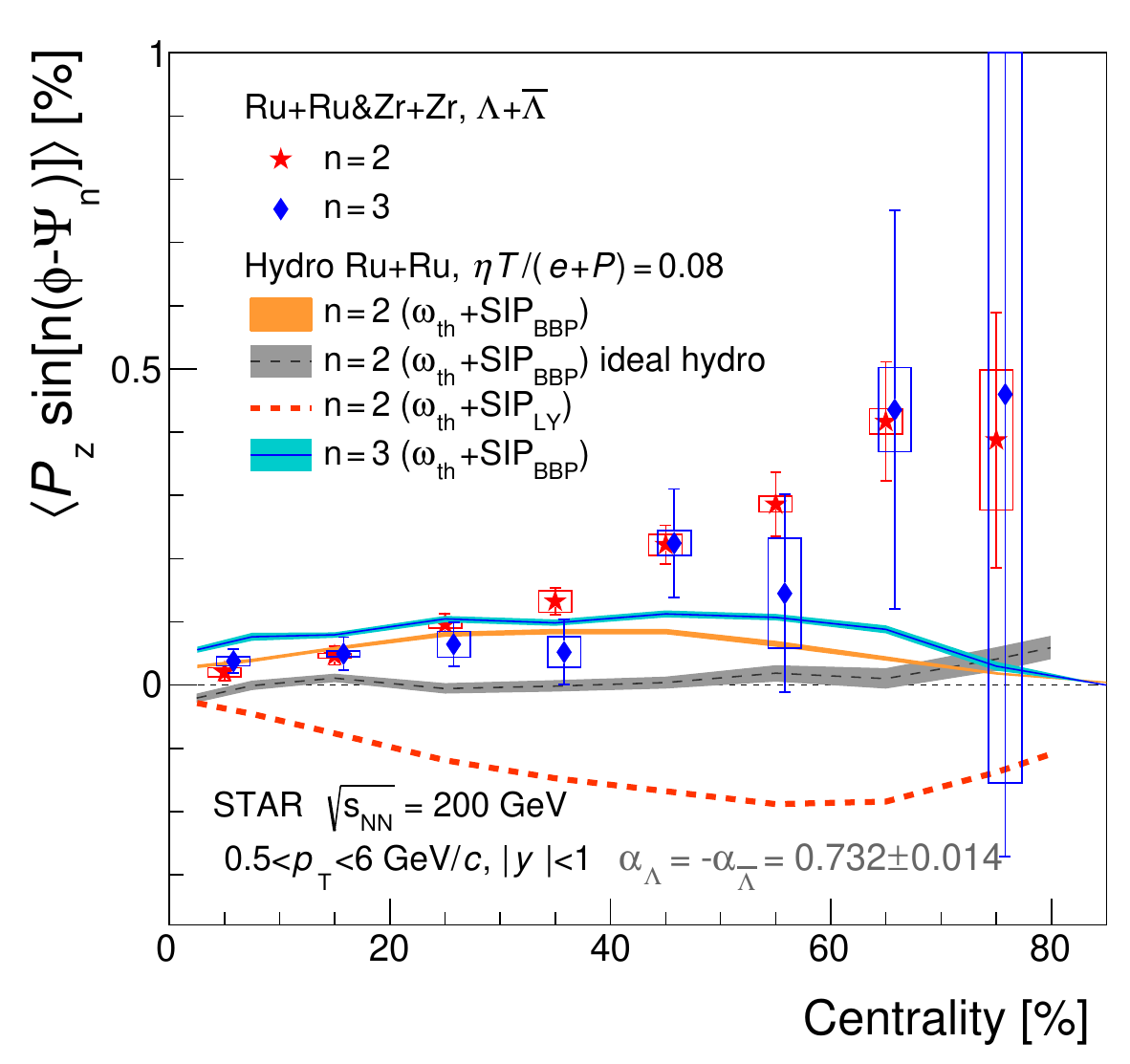}
\caption{Centrality dependence of the second- and the third-order Fourier sine coefficients of $\Lambda$/$\bar\Lambda$ polarization along the beam direction in isobar Ru+Ru and Zr+Zr collisions at $\sqrt s$ 200 GeV. 
Solid bands show calculations from hydrodynamic model.
Figure taken from \cite{prl_local_isobar_2023}.
}
\label{fig:PzvsCent}
\end{center}
\end{figure}

Beyond global polarization along the system’s orbital angular momentum, recent studies have uncovered more intricate vorticity structure in the QGP. The measurement of flow-driven local polarization ($P_z$) revealed a distinct sine modulation with the azimuthal angle~\cite{prl_local_auau200_2019, prl_local_isobar_2023}, suggesting the presence of local vortical and shear fields. 
Recently, the local polarization of $\Lambda$+$\bar\Lambda$ along the beam direction has been measured relative to the second and third harmonic event planes in isobar Ru+Ru and Zr+Zr collisions at $\sqrt s$ = 200 GeV at STAR, which provides the first evidence of local polarization by the triangular flow originating from the initial state geometry~\cite{prl_local_isobar_2023}.
As seen from Fig.~\ref {fig:PzvsCent}, the amplitudes of the sine modulation for the second and third harmonic are comparable, increasing from central to peripheral collisions. 
The azimuthal angle dependence of the polarization follows the vorticity pattern expected due to elliptic and triangular flow, but qualitatively disagree with most hydrodynamic model calculations based on thermal vorticity and shear induced contributions~\cite{prl_local_isobar_2023}.

Another striking discovery is the observation of large global spin alignment for vector mesons, including $\phi$, $K^{*0}$ and $J/{\ensuremath{\psi}}$~\cite{nature_spin_alignment_2023,STAR:2008lcm,ALICE:2022dyy}. The spin-density matrix element $\rho_{00}$ deviates substantially from the unpolarized value of $1/3$, with magnitudes not yet reproduced by hydrodynamic or partonic spin-transport models. These results may hint at contributions from strong electromagnetic fields or spin--spin correlations~\cite{Lv:2024uev,Chen:2024afy,Sheng:2025puj}. The measurement has recently been extended to heavy-flavor mesons such as $J/\psi$ and $D^{+}$~\cite{ALICE:2022dyy}, opening a new avenue to probe charm-quark spin dynamics in the QGP.

Despite these advances, key puzzles remain. The so-called ``spin sign puzzle,'' concerning the sign of the local polarization $P_z$, may require the inclusion of shear-induced polarization and non-equilibrium effects in theoretical descriptions~\cite{Fu:2021pok,Arslan:2024dwi}. Future efforts aim at high-precision differential measurements, searches for possible spin Hall effects, chiral magnetic effect, and systematic studies of spin--spin correlations etc~\cite{Liu:2020dxg,STAR:2021mii,Guo:2025wry,Lv:2024uev,STAR:2025vhs,Sheng:2025puj}. Together, these developments establish spin as a new and sensitive probe of the vortical and quantum structure of the QGP.

\section{Summary}
This review presents a comprehensive summary of selected highlights from the RHIC-STAR experiment with key contributions from STAR-China group, with the objective of characterizing the properties of deconfined partonic matter using multiple experimental probes. These include collective flow phenomena, thermal dileptons and UPC, heavy-flavor and jet production, antimatter hypernucleus formation, nuclear structure studies, and spin polarization  etc. 

The elliptic flow measurements in Au+Au collisions at top energy reveal a unified picture of the QGP as a strongly coupled, nearly perfect fluid,  with strong evidence that collectivity is established at the partonic level prior to hadronization. 
And the collisions of small systems have become a crucial probe of collective behavior and the conditions required for QGP formation.
Thermal dileptons provide a unique and penetrating thermometer for the QGP, and recent measurements provide the first multi-energy extraction of QGP temperatures at distinct evolutionary stages.
On the other side, heavy flavor quarks and energetic partons or jets are predominately produced in the initial hard scattering, and thus provide penetrating probes of the QGP. Recent measurements of bottomonium and charmonium states are consistent with the picture of sequential suppression of quarkonia. The underlying event study with jet production in p+p collisions enhances the precision of jet tomography as a probe for QGP properties. The global polarization measurements have been established as a unique probe of vortical structure of QGP. In addition, the novel nuclear shape-imaging studies from the collective response of debris in relativistic heavy-ion collisions, provide instantaneous “snapshot” of the nuclear geometry. 

The aforementioned results by the STAR Collaboration collectively contribute to a deeper understanding of the QGP dynamics and the intrinsic structure of nuclei and nucleons.
Earlier this year, STAR concluded its remarkably successful 25-year operational period. Nevertheless, the high-statistics datasets from Au+Au and p+p collisions, acquired in recent years utilizing the upgraded STAR detector, are expected to produce exciting physics results in forthcoming years. These include, notably, investigations into the QCD critical point search through the Beam Energy Scan II program, as well as studies of spin physics in forward rapidity region, etc.  


\section*{Acknowledgements}
The authors are grateful for the STAR Collaboration at RHIC.
This work is supported in part by the National Key Research and Development Program of China under Contract No. 2022YFA1604900, and the National Natural Science Foundation of China (Grant No.12575145).


\bibliography{reference.bib}

\end{document}